\UseRawInputEncoding
\documentclass[10pt,nofootinbib,preprint,showpacs,showkeys,floatfix,aps,a4paper]{revtex4-1}
\usepackage{amsmath,multirow,amssymb,mathtools}
\usepackage[T1]{fontenc}
\usepackage{verbatim}
\usepackage{float} 
\usepackage{amsfonts}
\usepackage{comment}
\usepackage{verbatim}
\usepackage{braket}
\usepackage{rotating}
\usepackage{dsfont}
\usepackage{xcolor}
\usepackage[justification=raggedright]{caption}
\usepackage{bm}
\usepackage{tikz}
\usetikzlibrary{matrix}
\usepackage{xpatch} %make hyperref work better with revtex 
\makeatletter
	\xpatchcmd{\@ssect@ltx}{\@xsect}{\edef\@currentlabelname{#8}\@xsect}{}{}% Patch \<section>*
	\xpatchcmd{\@sect@ltx}{\@xsect}{\edef\@currentlabelname{#8}\@xsect}{}{}% Patch \<section>
\makeatother
\usepackage{hyperref}

\usepackage[sort&compress]{natbib}
\setcitestyle{numbers}
\hypersetup{
    colorlinks=true,
    linkcolor=blue,
    filecolor=magenta,      
    urlcolor=cyan,
    pdftitle={Overleaf Example},
    pdfpagemode=FullScreen,
    }

\newcommand{\ownref}[2]{\hyperref[#2]{#1~\ref*{#2}}}% own ref text: \ownref{<text>}{<label key>}
\usepackage{cleveref}
\usepackage{xcolor}
\graphicspath{ {./Figures/} }
\begin{document}

\title{An Electro-Optical Trap for Polar Molecules}

\author{Bretislav Friedrich}

\email[On leave from Fritz-Haber-Institut der Max-Planck-Gesellschaft, Faradayweg 4-6, D-14195 Berlin, Germany. Email: \hspace{0.05cm}]{bretislav.friedrich@fhi-berlin.mpg.de}
\vspace{1cm}

\address{Department of Physics, Harvard University\\ 17 Oxford St, Cambridge, MA 02138, U.S.A.  \\
\text{and}\\
%\vspace{1cm}
Center for Astrophysics | Harvard \& Smithsonian,
Institute for Theoretical Atomic Molecular and Optical Physics (ITAMP), 60 Garden St, Cambridge, MA 02138, U.S.A.}

\date{\today}

\begin{abstract}
A detailed treatment of an electro-optical trap for polar molecules, realized by embedding an optical trap within a \emph{uniform} electrostatic field, is presented and the trap's  properties analyzed and discussed. The electro-optical trap offers significant advantages over an optical trap that include an increased trap depth and conversion of alignment of the trapped molecules to marked orientation. Tilting the polarization plane of the optical field with respect to the electrostatic field diminishes both the trap depth and orientation and lifts the degeneracy of the $\pm M$  states of the trapped molecules. These and other features of the electro-optical trap are examined in terms of the eigenproperties of the polar and polarizable molecules subject to the combined permanent and induced electric dipole interactions at play. 
\end{abstract}


\maketitle

\section{Introduction}

Although keenly anticipated almost three decades ago \cite{Maddox1995,KadorACh1995}, the heyday of optical trapping of molecules arrived only recently along with the techniques to laser-cool molecular translation down to the ultracold regime ($\le 1$ mK), see Refs. \cite{DiRosa_EPJD_2004,DeMille_Transversal_PRL_2009,DeMille_BeamSlowing_PRL_2012,Ye_2DMOT_PRL_2013,
DeMille_MOT_Nature_2014,Truppe_BeloeDoppler_NaturePhysics_2017,Doyle_RFMOT_PRL_2017,Doyle_Sisyphus_PRL2017,Doyle_Ye_3DMOT_YO_PRL_2018,Tarbutt_LaserCooling_PRL2019,Ye_SubDopplerCooling_PRX_2020,DoyleLaserCool_Polyatomic_NJP_2020,Doyle_SymTopLaserCool_Science_2020} as well as recent reviews \cite{Jesus_Book_2020,FitchTarbutt2021,FitchTarAAMOP2021}. However, optical traps had been loaded as early as 1998 with ultracold molecules produced by dimer formation from ultracold atoms in a MOT \cite{Knize_PRL_Cs2_1998} and in the 2000s by magneto-association \cite{Gabbanini_OptTrap_2004,Ni_Ye_Science2008,Grimm_RbCs_PRA_2009} or by photo-association \cite{Pillet_PRL_Cs2_2002,MarcassaPRA2011} of ultracold atoms.

Based on high-field-seeking states created by the purely attractive interaction of molecular polarizability with a far-off resonant optical field, optical traps have been coveted for their ability to trap ground-state molecules (as these are always high-field seeking) as well as their versatility (weak species dependence). The reliance of optical traps on a maximum of electric field strength in free space produced by focusing a laser beam circumvents limitations on molecular trapping imposed by Earnshaw's theorem for magnetic and other traps based on low-field seekers, see, e.g.,  Ref. \cite{PattersonChemPhysChem2016,DeMille_MagneticTrapping_PRL_2018}. 

Among the recently demonstrated advantages of optical traps (or of optical tweezers, their variant that makes use of tight, diffraction-limited focusing of the optical field) are long coherence times of the trapped samples \cite{KettNiDoylePRL2021}. These are key to such applications as searches for physics beyond the Standard Model \cite{Safronova2018} and quantum computing and quantum simulation \cite{Yelin2006,Quantum_Info4,BohnReyYeScience2017}. At the same time, optical traps are compatible with laser \cite{KettDoyleNatPhys2018} and sideband \cite{CaldTarPRR2020} cooling of the trapped molecules as well as with control of the molecules' mutual interactions \cite{Bohn_Shielding_2016,NiKettDoyleScience2021,Bloch_arXiv2022,Rempe_DipShielding_arXiv_2022} -- both critical to achieving quantum degeneracy in molecular systems \cite{Zoller2012}. The compatibility of optical traps also extends to optical imaging of the trapped molecules \cite{Doyle_Imaging_in_OptTrap_PRL_2018} as well as to optical cavities \cite{Stamper_Kurn_PRX_2021}. Last but not least, optical tweezers have played a central role as ``beakers'' for building molecules atom-by-atom via photo-association and in studying the detailed dynamics of the collisional processes involved \cite{NiScience2018,Ni_PRA2021}. 

In addition, an optical trap makes the trapped molecules directional -- aligned -- by virtue of the anisotropic interaction of the molecular polarizability with the polarization vector of the optical trapping field \cite{FriHerPRL1995,FriHerJPC1995}. If the trapped molecules are polar, their alignment (which corresponds to a double-headed arrow) by the optical field can be converted to orientation (corresponding to a single-headed arrow) by superimposing an electrostatic field \cite{FriHerJCP1999,FriHerJPC1999,FriRSC2021}. Polar molecules confined in tweezer arrays \cite{LukinScience2016,Ni_Doyle_Science2019} entangled via the electric-dipole-dipole interaction between the molecules have been envisioned as platforms for quantum computing with the oriented molecular states serving as qubits \cite{Quantum_Info0,Quantum_Info4}.

Herein, we provide a detailed quantum treatment of the optical trap for molecules and extend it to the case when the optical trap is embedded within a \emph{uniform} electrostatic field. For polar molecules, the resulting \emph{electro-optical trap} offers significant advantages over trapping by an optical field alone that include increased trap depth, apart from orienting the trapped molecules. The quantum effects involved in increasing the effective inhomogeneity of the optical field (due to a focused Gaussian laser beam) by a uniform electrostatic field as well as the enhancement of molecular orientation due to the electrostatic field by the optical field have been of interest in their own right \cite{SchmiFriJCP2014,SchmiFriPRA2015,SchatzFriBeckSchmiPRA2018} and will be laid out and explored below in the context of molecular trapping.

This paper is structured as follows: In Section \ref{sec:potentials}, the permanent and induced dipole interactions are introduced and the corresponding molecular potentials derived for a far-off resonant Gaussian optical field and a uniform electrostatic field. Section \ref{sec:eigenproblem} treats the eigenproblem of a polar and polarizable rigid rotor subject to the combined permanent and induced electric dipole potentials. Subsections \ref{sec:eigenprop_induced} and \ref{sec:combined} present, respectively, the eigenproperties due to the induced dipole potential alone and due to the combined permanent and induced dipole potentials. Included is a table of molecular constants for a selection of polar molecules together with their relation to the dimensionless parameters used to characterize  the strengths of the permanent and induced dipole interactions. 
Section \ref{sec:trap_pot} explores the properties of the optical trap (Subsection \ref{sec:opt}) and of the electro-optical trap (Subsection \ref{sec:el_opt}) as given by the parametric dependence of the eigenvalues of the corresponding Hamiltonians on the spatial coordinates. Subsection \ref{sec:harm} deals with the electro-optical trap in the harmonic approximation.  Finally,  trapping in perpendicular optical and electrostatic fields is treated in Subsection \ref{sec:perp}. Section \ref{sec:concl} provides a summary of the present work and extolls its most promising applications. Appendices \ref{sec:appendix1} and \ref{sec:appendix2} provide details of the derivations and calculations presented in this paper.

\section{Permanent and Induced Electric Dipole Interactions}
\label{sec:potentials}

\begin{figure}
%[htbp]
\centering
\includegraphics[width=3cm]{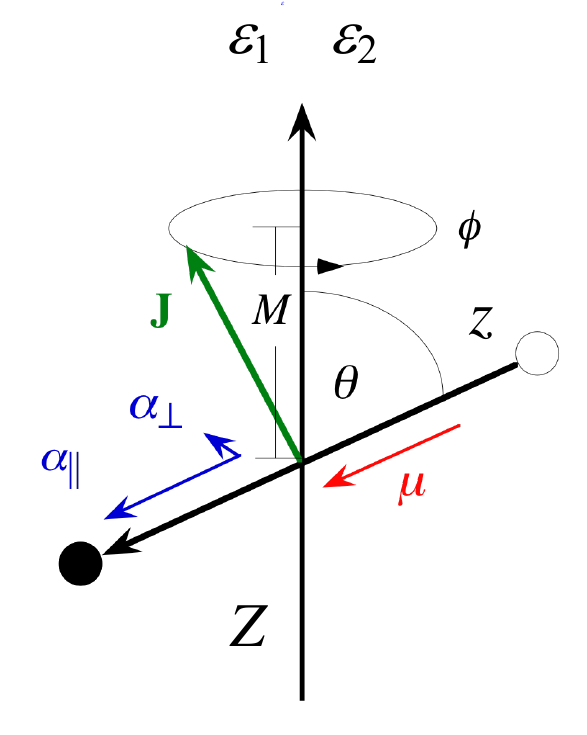}
\caption{\label{fig:Molecule} \footnotesize Electrostatic field $\boldsymbol{\varepsilon}_1$ and optical field $\boldsymbol{\varepsilon}_2$ of a linearly polarized laser acting jointly on the molecular dipole moment $\mu$ and the parallel, $\alpha _{\parallel}$, and perpendicular, $\alpha _{\bot }$, components of the molecular polarizability. Also shown are the polar angle $\theta$ between the space-fixed axis $Z$, as defined by the common direction of the field vectors, and the body-fixed (molecular) axis $z$, as well as the projection $M$ of the angular momentum $\bf J$  on the $Z$ axis together with the uniformly distributed azimuthal angle $\phi$ of ${\bf J} \perp z$ about $Z$.  See also text and Appendix \ref{sec:appendix1}.}
\end{figure}

We consider a polar and polarizable linear rotor subject to a combination of an electrostatic field $\boldsymbol{\varepsilon}_1=(\varepsilon_{1,X},\varepsilon_{1,Y},\varepsilon_{1,Z})$ and a far-off resonant/nonresonant optical field $\boldsymbol{\varepsilon}_2=(\varepsilon_{2,X},\varepsilon_{2,Y},\varepsilon_{2,Z})$ and assume that the fields only possess non-zero components $\varepsilon_{1,Z}\equiv \varepsilon_1$ and $\varepsilon_{2,Z}\equiv \varepsilon_2$ along the $Z$ axis of the space-fixed frame $X,Y,Z$, see Fig. \ref{fig:Molecule}.  For a molecule whose permanent dipole moment $\boldsymbol{\mu}=(\mu_x,\mu_y,\mu_z)$ has only a non-vanishing $z$-component $\mu_z\equiv \mu$ in the body-fixed frame $x,y,z$ and whose polarizability tensor, $\boldsymbol{\alpha}$, 
\begin{equation}
\label{principal}
\boldsymbol\alpha=\left(
\begin{matrix}
    \alpha_{xx} & 0 & 0\\
    0 & \alpha_{yy} & 0\\
    0 & 0 & \alpha_{zz}
  \end{matrix}\right)
\end{equation}
has principal components $\alpha_{xx}=\alpha_{yy}\equiv \alpha_{\perp}<\alpha_{zz}\equiv \alpha_{\parallel}$ in that frame, 
the permanent and induced dipole potentials are given, respectively, by
\begin{equation}
\label{Vmu}
V_{\mu}(\theta)=-\mu (\varepsilon_1+\varepsilon_2) \cos\theta
\end{equation}
and
\begin{equation}
\label{Valpha3}
V_{\alpha}(\theta)=-(\varepsilon_1+\varepsilon_2)^2\left(\Delta\alpha \cos^2\theta+\alpha_{\perp}\right)
\end{equation}
where $\theta$ is the polar angle between the body- and space-fixed axes $z$ and $Z$ and $\Delta\alpha \equiv \alpha_{\parallel}-\alpha_{\perp}$, see also Appendix \ref{sec:appendix1}. In what follows, we assume that  $\varepsilon_2$ is due to a far-off resonant electromagnetic wave plane-polarized along the space-fixed axis $Z$,  
\begin{equation}
\varepsilon_{2}=\varepsilon_0 \cos(2\pi \nu t)
\end{equation}
where $\varepsilon_0$ is the wave's amplitude and $\nu$ its frequency. Then, for nonresonant frequencies much greater than the reciprocal of the time, $\tau$, the field is on, $\nu \gg \tau^{-1}$, averaging over $\tau$ quenches the permanent dipole interaction with $\varepsilon_2$,
\begin{equation}
\label{Vmuave}
\overline{V}_{\mu}(\theta)=\frac{1}{\tau} \int_0^{\tau} V_{\mu}dt=-\mu \varepsilon_1 \cos\theta
\end{equation}
and converts $\varepsilon_2^2$ in the polarizability term to $\frac{1}{2}\varepsilon_0^2$,
\begin{equation}
\label{Valphave}
\overline{V}_{\alpha}(\theta)=\frac{1}{\tau} \int_0^{\tau} V_{\alpha}dt=-\frac{1}{2}\varepsilon_0^2\Delta \alpha \cos^2 \theta-\frac{1}{2}\varepsilon_0^2\alpha_{\perp}
\end{equation}
where we made use, in addition, of  the disparity in the magnitudes of  the electrostatic and optical fields, $\varepsilon_1 \ll \varepsilon_2$.
 
A Gaussian laser beam \cite{Erikson_Beams_1994} of power $P$ plane-polarized along the $Z$ axis propagating along the $X$-axis and focused to a waist $w_0$ has an intensity $I(X,Z)$
 \begin{equation}
 \label{intensity}
 I(X,Z)=\frac{2P}{\pi w^2(X)}\exp\left[-\frac{2Z^2}{w^2(X)} \right]=\frac{I_0}{1+\left (\frac{X}{X_R}\right)^2}\exp\left[-\frac{2Z^2}{w^2(X)}\right]
\end{equation} 
 with 
\begin{equation}
 w(X)=w_0\left[1+\left(\frac{X}{X_{\mathrm{R}}}\right)^2\right]^{1/2}
 \end{equation}
where
\begin{equation}
\label{Rayleigh}
X_{\mathrm{R}}\equiv \frac{\pi \nu w_0^2}{c}
\end{equation}
is the Rayleigh length and $I_0\equiv I(0,0)$ the maximum beam intensity. The Gaussian beam gives rise to an electric field amplitude along the $Z$ axis
\begin{equation}
 \varepsilon_0=\left(\frac{2I}{\epsilon_0 c}\right)^{1/2}
 \end{equation}
where $\epsilon_0$ is the electric permitivity and $c$ the speed of light in vacuum. As a result, the induced-dipole, optical potential, $\overline{V}_{\alpha}(\theta)$, becomes
\begin{equation}
\label{VindI}
\overline{V}_{\alpha}(\theta)=-\frac{I}{\epsilon_0 c}\left(\Delta \alpha \cos^2 \theta+\alpha_{\perp}\right)
\end{equation}

The polarizability components $\alpha_{\parallel}$ and $\alpha_{\bot}$ depend on the frequency $\nu$ of the laser field. A detailed treatment of this dependence and more has been given in Refs.\cite{CaldTarPRR2020,FitchTarAAMOP2021}. Static polarizabilities, such as those listed in Table \ref{table:parameters}, approximate well the dynamic ones at low-enough laser frequencies, cf. Eq. (76) of Ref. \cite{FitchTarAAMOP2021}. The theory of electric multipole moments has been reviewed in Ref. \cite{Kalugina_Book_2017}. We note that for tight focusing of the optical field, linear polarization may not be achievable \cite{Erikson_Beams_1994}, in which case an interaction with additional components of the polarizability tensor of the molecule beyond those included in Eq. (\ref{Valpha3}) has to be considered. 

\begin{table}[h]

\centering
\caption{\footnotesize Molecular parameters of representative linear molecules. Also given are the values of the permanent, $\eta$, and induced-dipole, $\Delta\zeta$, interaction parameters at choice values of the electrostatic field $\varepsilon$ and laser intensity $I$. Note that $\eta=0.0168\, \mu$ [Debye] $\varepsilon_1$ [kV/cm]/$B$  [cm$^{-1}$]$=5.05\times 10^2\mu$ [Debye] $\varepsilon_1$ [kV/cm]/$B$  [MHz], $\zeta_{\parallel,\perp}=1.05 \times 10^{-11} I$ [W/cm$^2$] $\alpha_{\parallel,\perp}$ [\AA$^3$]$/B$  [cm$^{-1}$] =$\zeta_{\parallel,\perp}=3.15 \times 10^{-7} I$ [W/cm$^2$] $\alpha_{\parallel,\perp}$ [\AA$^3$]$/B$  [MHz]. In particular, $\varepsilon_0$ [kV/cm]$=1.941\times 10^{-2} I$ [W/cm$^2$]. For a laser power $P=0.4$ W at $\lambda = c/\nu = 780$ nm focused to a waist $w_0=2\lambda$, the laser intensity $I \approx 1\times 10^{7}$ W/cm$^2$; for a polarizability anisotropy $\Delta\alpha = 1$ \AA$^3$ and rotational constant $B=1$ cm$^{-1}$, the dimensionless interaction parameter $\Delta \zeta \approx 10^{-4}$. The conversion factor for the rotational period, $\tau_{\mathrm{r}}=\frac{\pi \hbar}{B}$, is $\tau_{\mathrm{r}}$ [ps] =$16.65/B$ [cm$^{-1}$]=$4.98\times 10^5/B$ [MHz].  The Table lists only static polarizabilities, with estimates based on bond polarizabilities in parentheses \cite{GreenBible} as well as Refs. \cite{DykstraJACS1995,HohmJMS2013}.}

\label{table:parameters}

\begin{tabular}{| l | c | c | c | c | c | c |}

\hline 
\hline
Molecule & $B$ [cm$^{-1}$] & $\mu$ [D] & $\eta$ @  $10$ kV/cm  & $\Delta \alpha$ [\AA$^3$] &  $\Delta\zeta$ @  $I =1\times 10^{7}$ W/cm$^2$ & $\tau_{\mathrm{r}}$ [ps]  \\[5pt] \hline

CsF(X$^1\Sigma$) & 0.1843  &  7.87 &  7.17 & (3.0) & (0.0163) & 90.33  \\[5pt]

ICN(X$^1\Sigma$) &  0.1075  &  3.72 &  5.81 & (7.0) & (0.0651) & 154.87  \\[5pt]

LiCs(X$^1\Sigma$) &  0.188 &  5.52 & 4.93  & 49.5 & 0.2633 & 88.56  \\[5pt]

NaK(X$^1\Sigma$) &  0.091 &  2.76 & 5.10  & 39.5 & 0.4341 & 182.95  \\[5pt]

KCs(X$^1\Sigma$) &  0.033 &  1.92 & 9.77  & 64.6 & 1.9576 & 504.51  \\[5pt]

RbCs(X$^1\Sigma$) &  0.016 &  1.27 & 13.34  & 72.8 & 4.5500 & 1040.54  \\[5pt]

ICl(X$^1\Sigma$) &  0.1142  &  1.24  &  1.82 & (9.0) & (0.0788) & 145.79  \\[5pt]

CO(A$^3\Pi$) &  1.681  &  1.37    &  0.14  & (1.5) & (0.0009) & 9.90  \\[5pt]

OCS(X$^1\Sigma$) &  0.2039  &  0.709 &  0.58 & 4.1 & 0.0201 & 81.65  \\[5pt]

KRb(X$^1\Sigma$) &  0.032 &  0.76 & 3.99  & 54.1 & 1.6906 & 520.27  \\[5pt]

LiNa(X$^1\Sigma$) &  0.38 &  0.566 & 0.25  & 24.7 & 0.0650 & 43.81  \\[5pt]

NO(X$^2\Pi$) &  1.703  &  0.16  & 0.016  & 2.8 & 0.0016 & 9.78  \\[5pt]

CO(X$^1\Sigma$) &  1.931  &  0.10 &  0.009 & 1.0 & 0.0005 & 8.62 \\[5pt]

HD(X$^1\Sigma$) &  45.644  &  5$\times 10^{-4}$ & $2\times10^{-6}$  & 0.305 & $6.7\times 10^{-6}$ & 0.36 \\[5pt]
 
\hline
\hline
  
\end{tabular}
\end{table}

\section{The eigenproblem for a polar and polarizable rigid-rotor molecule subject to combined permanent and induced electric dipole interactions}
\label{sec:eigenproblem}

The Hamiltonian of a $^{1}\Sigma $ rigid-rotor molecule subject to the combined permanent and induced dipole potentials of Eqs. (\ref{Vmuave}) and (\ref{VindI}) is given by
\begin{equation}
H=B\mathbf{J}^{2}+\overline{V}_{\mu }+\overline{V}_{\alpha }  \label{hamiltonian}
\end{equation}
where $\mathbf{J}^{2}$ is the operator of the angular momentum squared and $B$ is the rotational constant \cite{FriHerPRL1995,FriHerJPC1995,FriRSC2021}. By dividing through $B$, the Hamiltonian becomes dimensionless, 
\begin{equation}
\frac{H}{B}=\mathbf{J}^{2}+\frac{\overline{V}_{\mu}}{B}+\frac{\overline{V}_{\alpha}}{B}  \label{redham}
\end{equation}
In particular, the dimensionless potentials become
\begin{equation}
\label{Vperm}
\frac{\overline{V}_{\mu}(\theta)}{B}=-\eta \cos \theta
\end{equation}
and
\begin{equation}
\label{Vind}
\frac{\overline{V}_{\alpha}(\theta)}{B}=-\Delta \zeta \cos^2 \theta-\zeta_{\bot}
\end{equation}
where
\begin{eqnarray}
\label{param}
\eta \equiv \frac{\mu \varepsilon_1}{B}  \hspace{1cm}
\zeta_{\parallel,\bot}\equiv \frac{I}{\epsilon_0 c B}\alpha_{\parallel,\bot}  \hspace{0.5cm}  \hspace{0.5cm} \Delta \zeta\equiv \zeta_{\parallel}-\zeta_{\bot}
\end{eqnarray}
are dimensionless parameters that characterize the strengths of the permanent and induced-dipole (polarizability) interactions. 

The eigenenergies, $E_{\tilde{J},|M|}/B$, and eigenfunctions, $\ket{\tilde{J},|M|;\eta,\Delta\zeta}$, obtained from the Schr\"{o}dinger equation pertaining to the dimensionless Hamiltonian (\ref{redham})
\begin{equation}
\frac{H}{B}\ket{\tilde{J},|M|;\eta,\Delta\zeta}=\frac{E_{\tilde{J},|M|}}{B}\ket{\tilde{J},|M|;\eta,\Delta\zeta}
\label{redSE}
\end{equation}
are arbitrarily ``transferrable'' for given values of the interaction parameters from one molecular species to another. Table \ref{table:parameters} lists the molecular parameters for a set of representative linear molecules as well as the corresponding values of the dimensionless parameters $\eta$ and $\Delta\zeta$ for choice values of the strength of the electrostatic field and of the laser intensity. Also included in Table \ref{table:parameters} are the requisite conversion factors.

The eigenproperties of Hamiltonian (\ref{redham}) can be obtained by expanding its eigenfunctions in the free-rotor basis set $\ket{J,|M|}$
\begin{equation}
\label{wf}
\ket{\tilde{J},|M|;\eta,\Delta\zeta}=\sum^{J_{\mathrm{max}}}_{J=|M|} c^{\tilde{J},|M|}_J(\eta,\Delta \zeta) \ket{J,|M|} 
\end{equation}
and diagonalizing the resulting Hamiltonian matrix truncated at $J=J_{\mathrm{max}}$. The matrix elements $\langle J',M'|H|J,M\rangle$ are listed in Appendix \ref{sec:appendix2}. 

The wavefunctions $\ket{\tilde{J},|M|;\eta,\Delta\zeta}$ are thus recognized as coherent linear superpositions, or \emph{hybrids}, of the field-free rotor states $\ket{J,|M|}$ for a fixed value of the good quantum number $|M|$ and for a range of values of $J$, which is, alas, not a good quantum number. 
Nevertheless, the states created by the combined interaction can be labeled by $|M|$ and the nominal value, $\tilde{J}$, of the angular momentum quantum number of the free-rotor state $\ket{J,|M|}$ with which they adiabatically correlate, $\ket{\tilde{J},|M|;\eta=0,\Delta \zeta=0}\rightarrow \ket{J,|M|}$. The hybridization coefficients, $c^{\tilde{J},|M|}_J(\eta,\Delta \zeta)$, depend, for a given hybrid state $\ket{\tilde{J},|M|;\eta,\Delta\zeta}$, solely on the interaction parameters $\eta$ and $\Delta \zeta$. Since the sense of rotation of the molecular dipole makes no difference in the combined collinear electric fields, only $|M|$, the magnitude of $M$, matters. The hybrid states are also referred to as \emph{pendular states}, a term emphasizing that the axis of molecules in these state can no longer rotate through $2\pi$ but rather librates within a limited angular range $<2\pi$.  

In practice, the number, $J_{\mathrm{max}}$, of $J$'s in the ground-state hybrid wavefunction is on the order of the interaction parameter, i.e., if the eigenproperties are to be evaluated with an accuracy sufficient for most applications. Generally, the higher the $\tilde{J}$ of a given state, the fewer rotational basis states are drawn into its hybrid wavefunction at a given value of $\Delta\zeta$. This is because of the $J(J+1)$ rotational energy ladder and hence the increasing separation of the rotational basis states that make up the hybrid. That there is no hybridization of the angular momentum projection quantum number $M$ has to do with the cylindrical symmetry of the problem about the two collinear electric field vectors $\boldsymbol{\varepsilon}_1$ and $\boldsymbol{\varepsilon}_2$. Once this symmetry is broken, i.e., if the field vectors are tilted, $M$ ceases to be a good quantum number and states with different $M$'s are drawn into the hybrid wavefunction as well. Moreover, the $\pm M$ degeneracy of the energy levels is lifted \cite{FriHerJCP1999,FriHerJPC1999,HartFriJCP2008}.

Note that in the absence of an anisotropic polarizability, i.e., for $\Delta \alpha=0$, the  $\Delta \alpha \cos^2\theta$ term vanishes, thereby precluding hybridization of rotational states by the induced dipole interaction. Likewise, the absence of the body-fixed permanent dipole moment, i.e., for $\mu=0$, as is the case for nonpolar molecules, would preclude hybridization of the rotational states by the permanent dipole interaction.

\subsection{Eigenproperties due to an induced dipole potential alone}
\label{sec:eigenprop_induced}

\begin{figure}
%[htbp]
\centering
\includegraphics[width=14cm]{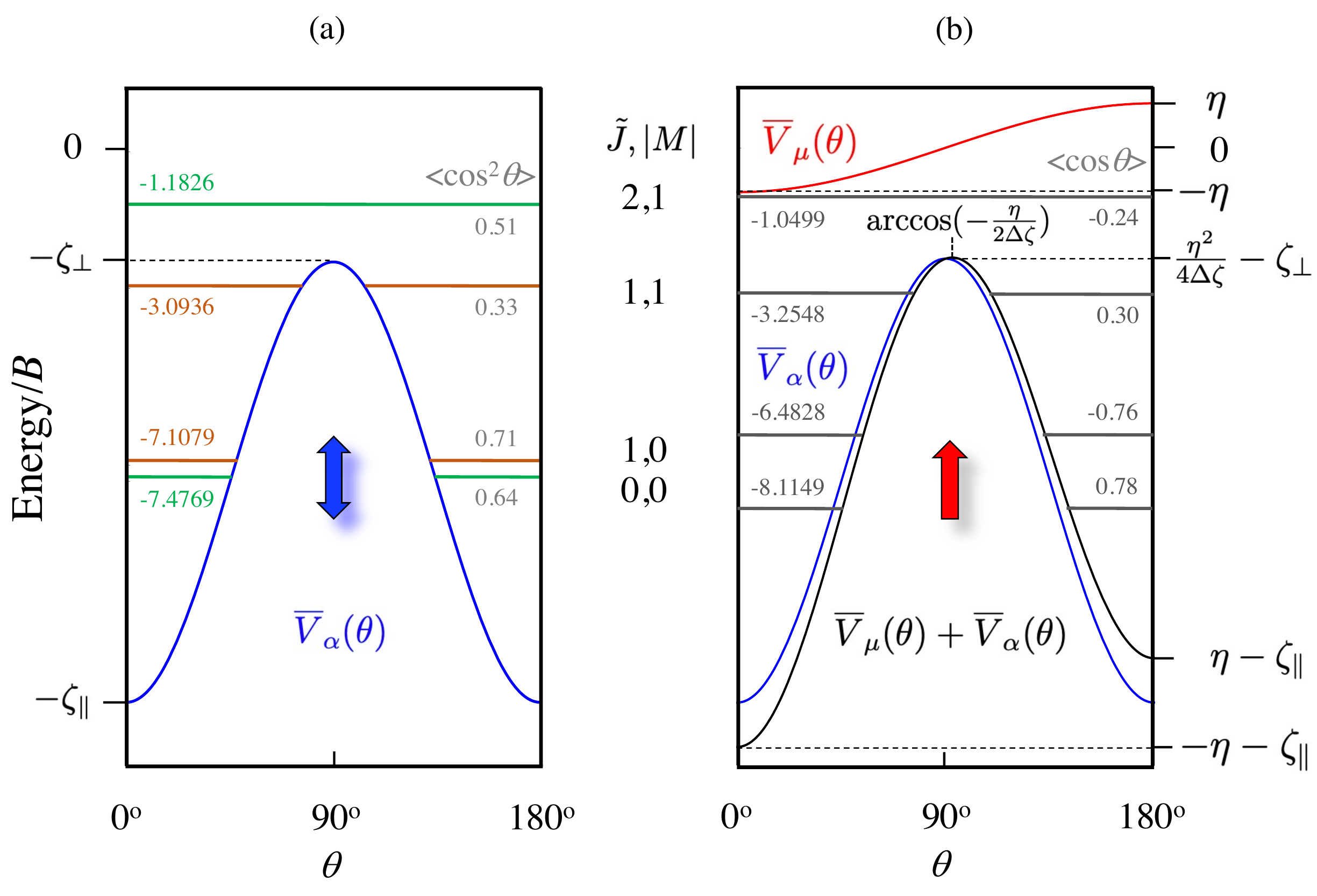}
\caption{\label{fig:Comparison_Opt_Comb_1_10} \footnotesize (a) Induced-dipole potential $\overline{V}_{\alpha}(\theta)$, Eq. (\ref{Vind}), as a function of the polar angle $\theta$. Note that $\overline{V}_{\alpha}(\theta)$ is a double-well potential with equivalent minima, $-\zeta_{\parallel}$, at $\theta=0^{\circ}$ and $180^{\circ}$ and a maximum, $-\zeta_{\bot}$, at $\theta=90^{\circ}$. The potential (in blue) has been drawn and the energy levels calculated for $\Delta\zeta=10$ and $\zeta_{\bot}=2.5$, which implies $\zeta_{\parallel}=12.5$. At these values of $\Delta\zeta$ and $\zeta_{\bot}$, only one tunneling doublet comprised of the $\ket{0,0;10}$ and $\ket{1,0;10}$ states is created by the double-well potential. Note that while the lower member, $\ket{1,1;10}$, of the next tunneling doublet is bound by the $\overline{V}_{\alpha}(\theta)$ potential, its upper member, $\ket{2,1;10}$, is not. Generally, the tunneling splitting decreases for more deeply bound doublets. Also note that the members of a given tunneling doublet have same $|M|$ but opposite parity (levels with $p=+1$ are shown in green, levels with $p=-1$ are shown in ochre). The $\zeta_{\perp}$ term shifts the potential -- and the energy levels it binds -- uniformly along the energy axis. The calculated energy levels bound by the induced-dipole potential are also shown along with the numerical values of their eigenenergies (left) and alignment cosines (right). The blue double-headed arrow indicates that the states created by the induced-dipole potential are aligned but not oriented. (b) Combined permanent and induced dipole potential $\overline{V}(\theta)=\overline{V}_{\mu}(\theta)+\overline{V}_{\alpha}(\theta)$ (black) of form A drawn for $\eta=1$, $\Delta\zeta=10$, and $\zeta_{\bot}=2.5$. The permanent dipole potential  $\overline{V}_{\mu}(\theta)$ and the induced dipole potential $\overline{V}_{\alpha}(\theta)$ are shown in red and blue, respectively. The calculated energy levels bound by the combined potential are also shown along with the numerical values of their eigenenergies (left) and orientation cosines (right). See also text.}
\end{figure}

A key feature of the induced-dipole interaction is that it couples free-rotor states whose $J$'s are either the same or differ by $\pm2$. As a result, the $\ket{\tilde{J},|M|;\Delta \zeta}$ states have definite parity, $p=(-1)^{\tilde{J}}$. 
\begin{figure}
%[htbp]
\centering
\includegraphics[width=14cm]{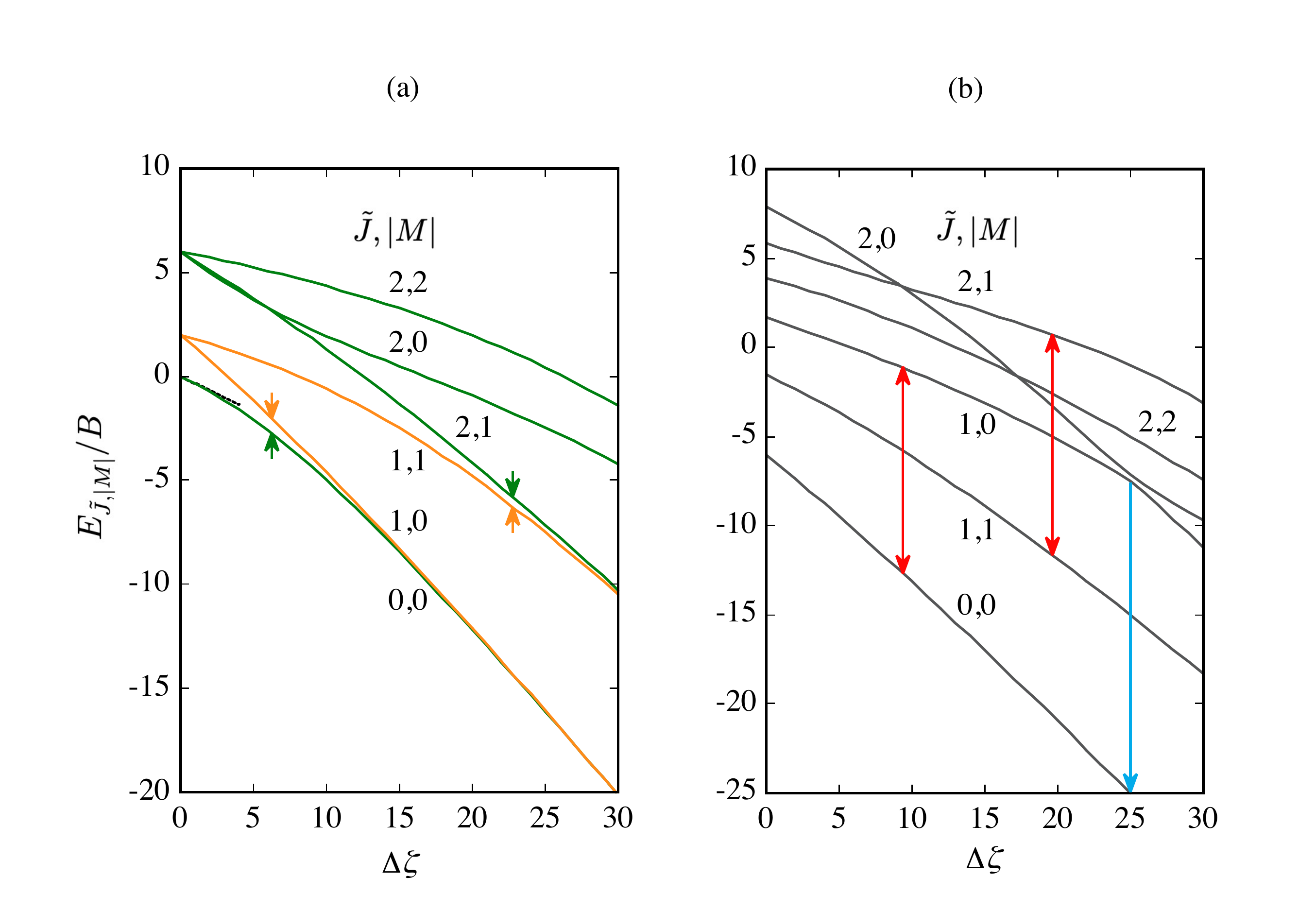}
\caption{\label{fig:Comparison_Eigenenergies_0_10} \footnotesize (a) 
The dependence of the eigenenergies $E_{\tilde{J},|M|}/B$ on the induced-dipole interaction parameter $\Delta\zeta$ for the lowest six initial rotational states of a linear polar molecule in the absence of an electrostatic field ($\eta=0$). 
Even-parity levels are shown in  green, odd-parity levels in ochre with pairs of similarly color-coded vertical arrows indicating the tunneling splitting of the tunneling doublets comprised of the $\ket{\tilde{J}=0,|M|=0;\Delta\eta}$ \& $\ket{\tilde{J}=1,|M|=0;\Delta\eta}$ and
$\ket{\tilde{J}=1,|M|=1;\Delta\eta}$ \&  $\ket{\tilde{J}=2,|M|=1;\Delta\eta}$ states. Cf. also panel (a) of Fig. \ref{fig:Comparison_Opt_Comb_1_10}. The black dotted curve shows the dependence of the ground-state eigenenergy in the low-field limit, cf. Table \ref{table:eigenproperties1} and text. Panel (b) displays the dependence of the eigenenergies on $\Delta\zeta$  in the presence of a superimposed electrostatic field that gives rise to a permanent dipole interaction with $\eta=10$. 
The red vertical arrows illustrate how the splitting of the tunneling doublets of panel (a) has been altered/enhanced by the superimposed electrostatic field.
The blue arrow shows the position, $\Delta\zeta=25$, of the avoided crossing between the $\ket{2,0;\eta=10,\Delta\zeta=25}$ and $\ket{1,0;\eta=10,\Delta\zeta=25}$ states, corresponding to the topological index $k=1$, see text.}
\end{figure}

The double-well nature of the induced dipole potential, Eq. (\ref{Vind}), causes all states bound by it to occur as doublets split by tunneling through the equatorial barrier, see panel (a) of Fig. \ref{fig:Comparison_Opt_Comb_1_10}.
The members of any given tunneling doublet have the same $|M|$ but opposite parities. 

The dependence of the eigenenergies of the six lowest pendular states on $\Delta \zeta$ is shown in panel (a) of Fig. \ref{fig:Comparison_Eigenenergies_0_10}. The tunneling splitting scales as $\propto \exp(-\Delta\zeta^{1/2})$ which means that the members of a given tunneling doublet can be drawn arbitrarily close to one another by boosting  $\Delta \zeta$, cf. Ref.\cite{FriRSC2021}. This is exemplified in the figure by the $\ket{\tilde{J}=0,|M|=0;\Delta\eta}$ \& $\ket{\tilde{J}=1,|M|=0;\Delta\eta}$ and
$\ket{\tilde{J}=1,|M|=1;\Delta\eta}$ \&  $\ket{\tilde{J}=2,|M|=1;\Delta\eta}$ tunneling doublets. 

The states that are created by the induced-dipole interaction of Eq. (\ref{Valphave}) are aligned, i.e., they behave like double-headed arrows pointing along the space-fixed axis $Z$. A measure of the directionality/alignment of the states is the expectation value of the $\cos^2 \theta$ operator,
\begin{equation}
\label{alignmentcos}
\langle \cos^2 \theta \rangle_{\tilde{J},|M|}=\langle \tilde{J},|M|;\Delta\zeta|\cos^2 \theta |\tilde{J},|M|;\Delta\zeta\rangle
\end{equation}
termed the \emph{alignment cosine}. It can be evaluated for a given state either directly from the state's wavefunction or via the Hellmann-Feynman theorem, 
\begin{equation}
\label{H_F}
\langle \cos^2 \theta \rangle_{\tilde{J},|M|}=-\frac{\partial E_{\tilde{J},|M|}(\Delta\zeta)}{\partial \Delta\zeta}
\end{equation}
Since the induced-dipole interaction is purely attractive, all states created by it are high-field seeking, cf. also panel (a) of Fig. \ref{fig:Comparison_Eigenenergies_0_10}, making the alignment cosine positive.  However, a given state can still be \emph{right-} or \emph{wrong-way aligned}, depending on whether the induced dipole (the molecular axis $z$) points along or perpendicular to the aligning field vector (the space-fixed axis $Z$). 

The eigenenergies -- and alignment cosines -- in the low- and high-field limit have been obtained in analytic form \cite{FriHerJPC1995} and are listed in Tables \ref{table:eigenproperties1} and \ref{table:eigenproperties2}.

\begin{table}[h]
\centering
\caption{\footnotesize Limiting values of eigenenergy, $E_{\tilde{J},|M|}/B$, of a linear molecule subject to the optical potential of Eq. (\ref{Vind}). See text and Ref. \cite{FriHerJPC1995}.}
\vspace{.3cm}
\label{table:eigenproperties1}
\begin{tabular}{| l | c | }
\hline 
\hline
Limit & $E_{\tilde{J},|M|}/B$  \\[5pt] \hline

$\Delta\zeta \rightarrow 0$ & $\tilde{J}(\tilde{J}+1)-\frac{\Delta\zeta}{2}\left[1-\frac{(2|M|-1)(2|M|+1)}{(\tilde{J}-1)(2\tilde{J}+3)} \right]-\zeta_{\bot}$    \\[5pt]

$\Delta\zeta \rightarrow \infty$ & $\Delta\zeta +2\Delta\zeta^{1/2}(\tilde{J}+1)+\frac{|M|^2}{2}-\frac{\tilde{J}^2}{2}-\tilde{J}-1- \zeta_{\bot}$ \qquad $\mathrm{for}$ \, $(\tilde{J}-|M|)$ \, $\mathrm{even}$\\[5pt]

$\Delta\zeta \rightarrow \infty$ & $\Delta\zeta +2\Delta\zeta^{1/2}\tilde{J}+\frac{|M|^2}{2}-\frac{\tilde{J}^2}{2}-\frac{1}{2}- \zeta_{\bot}$ \hspace{1.4cm} \qquad $\mathrm{for}$ \, $(\tilde{J}-|M|)$ \, $\mathrm{odd}$
\\[5pt]

\hline
\hline
  
\end{tabular}
\end{table}

\begin{table}[h]
\centering
\caption{\footnotesize Limiting values of alignment, $\langle \cos^2 \theta \rangle_{\tilde{J},|M|}$, of a linear molecule subject to the optical potential of Eq. (\ref{Vind}). See text and Ref. \cite{FriHerJPC1995}.}
\vspace{.3cm}
\label{table:eigenproperties2}
\begin{tabular}{| l | c | }
\hline 
\hline
Limit & $\langle \cos^2 \theta \rangle_{\tilde{J},|M|}$ \\[5pt] \hline

$\Delta\zeta \rightarrow 0$ &   $\frac{1}{2}\left[ 1-\frac{(2|M|-1)(2|M|+1)}{(\tilde{J}-1)(2\tilde{J}+3)}\right]+\Delta\zeta \left[\frac{(\tilde{J}-|M|+1)(\tilde{J}-|M|+2)(\tilde{J}+|M|+1)(\tilde{J}+|M|+2)}{2(2\tilde{J}+1)(2\tilde{J}+3)^3(2\tilde{J}+5)} \right]$   \\[5pt]

$\Delta\zeta \rightarrow \infty$ &  $1-\frac{\tilde{J}+1}{\Delta\zeta^{1/2}}$ \qquad $\mathrm{for}$ \, $(\tilde{J}-|M|)$ \, $\mathrm{even}$\\[5pt]

$\Delta\zeta \rightarrow \infty$ 
&  $1-\frac{\tilde{J}}{\Delta\zeta^{1/2}}$ \qquad $\mathrm{for}$ \, $(\tilde{J}-|M|)$ \, $\mathrm{odd}$\\[5pt]

\hline
\hline
  
\end{tabular}
\end{table}

Given the small values of the interaction parameter $\Delta\zeta$ that can be attained at feasible cw laser intensities ($I \approx 10^7$ W/cm$^2$), cf. Table \ref{table:parameters} ($\Delta\zeta \le 1$ for most of the molecules listed), the low-field limit is the relevant one for optical traps. In which case, the ground-state energy and alignment are accurately given by
\begin{equation}
\label{gse}
E_{0,0}/B=-\frac{1}{3}\Delta\zeta-\zeta_{\bot}
\end{equation}
and
\begin{equation}
\label{gsa}
\langle \cos^2 \theta \rangle_{0,0}=\frac{1}{3}+\frac{2}{135}\Delta\zeta
\end{equation}
revealing that the reduced eigenenergy is on the order of the interaction parameter $\Delta\zeta$, see the black dashed curve in panel (a) of Fig. \ref{fig:Comparison_Eigenenergies_0_10}, and that the alignment is puny (the spatial distribution of the molecular axis is nearly isotropic). Moreover, in the low-field limit, the upper member of the corresponding tunneling doublet (with $\tilde{J}=1,|M|=0$) will not be bound by the optical potential. Notable exceptions to the above are the highly polarizable heavy rotors such as RbCs or KRb, cf. Table \ref{table:parameters}, for which $\Delta\zeta$ on the order of 10 would be achieved at $I\approx 10^7$ W/cm$^2$. The eigenenergies of the lowest states are well rendered by the analytic expressions of Table \ref{table:eigenproperties1} in the low-field limit up to about $\Delta\zeta \le 1$, cf. Fig. \ref{fig:Comparison_Eigenenergies_0_10}. For stronger interactions, the eigenenergy of the molecule subject to the optical potential of Eq. (\ref{Vind})  has to be calculated by diagonalizing the corresponding truncated Hamiltonian matrix.

\subsection{Eigenproperties due to combined permanent and induced dipole potentials}
\label{sec:combined}

\begin{figure}
%[htbp]
\centering
\includegraphics[width=14cm]{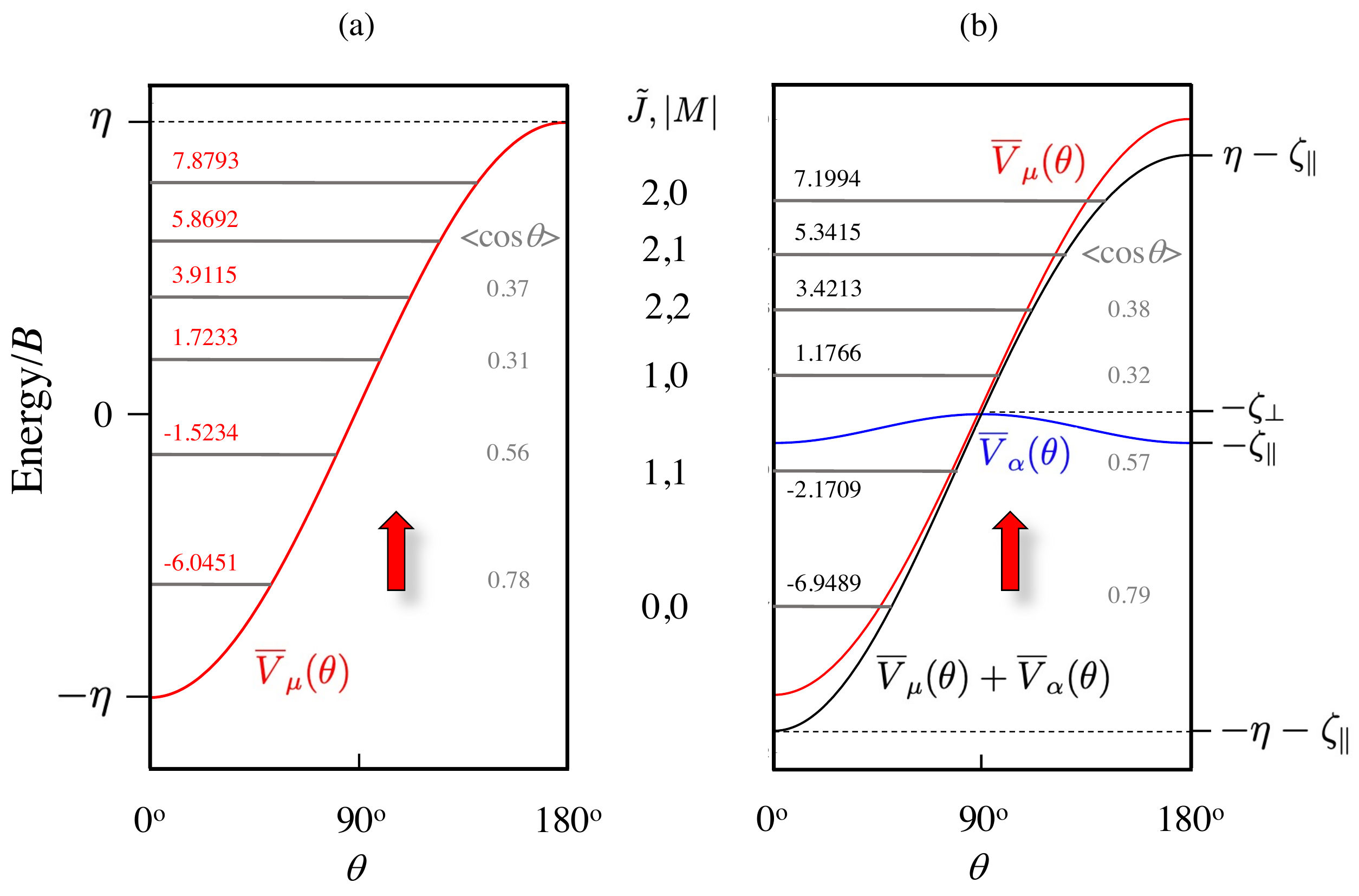}
\caption{\label{fig:Comparison_Electro_10_Opt_Comb_1_10} \footnotesize (a) Permanent dipole potential $\overline{V}_{\mu}(\theta)$ drawn for $\eta=10$ along with the calculated energy levels and their numerical values (left) and orientation cosine (right). The red vertical arrow indicates that the states created by the permanent dipole potential are oriented. 
(b) Combined permanent and induced dipole potential $\overline{V}(\theta)=\overline{V}_{\mu}(\theta)+\overline{V}_{\alpha}(\theta)$ (black) of Form B drawn for $\eta=10$, $\Delta\zeta=1$, and $\zeta_{\bot}=0.25$. The permanent dipole potential  $\overline{V}_{\mu}(\theta)$ and the induced dipole potential $\overline{V}_{\alpha}(\theta)$ are shown in red and blue, respectively. The calculated energy levels bound by the combined potential are also shown along with the numerical values of their eigenenergies (left) and orientation cosines (right).}
\end{figure}

Superimposing an electrostatic field onto the optical field changes dramatically the interaction potential and, consequently, the eigenstates of the polar and polarizable rotor. On the one hand, the permanent dipole interaction \emph{orients} the molecules. On the other, it changes the order of the energy levels: while for the pure induced-dipole interaction, the lowest energy state for a given $\tilde{J}$ has $|M|=0$, it is the ``stretched state,'' with $|M|=\tilde{J}$, that has the lowest energy for the pure permanent dipole interaction, cf. panels (a) of Figs. \ref{fig:Comparison_Opt_Comb_1_10} and \ref{fig:Comparison_Electro_10_Opt_Comb_1_10}. 

Oriented states behave like single-headed arrows pointing along the space-fixed axis $Z$. Their orientation is characterized by the expectation value of the $\cos \theta$ operator,
\begin{equation}
\label{alignmentcos}
\langle \cos \theta \rangle_{\tilde{J},|M|}=\langle \tilde{J},|M|;\eta,\Delta\zeta|\cos \theta |\tilde{J},|M|;\eta,\Delta\zeta\rangle
\end{equation}
termed the \emph{orientation cosine}. Like the alignment cosine, it can be evaluated for a given state either directly from the state's wavefunction or via the Hellmann-Feynman theorem, 
\begin{equation}
\label{H_F_2}
\langle \cos \theta \rangle_{\tilde{J},|M|}=-\frac{\partial E_{\tilde{J},|M|}(\eta;\Delta\zeta)}{\partial \eta}
\end{equation}

Depending on the relative magnitude of the permanent and induced dipole interaction parameters $\eta$ and $\Delta\zeta$, the combined permanent and induced dipole potential 
\begin{equation}
\overline{V}(\theta)\equiv \overline{V}_{\mu}(\theta)+\overline{V}_{\alpha}(\theta)
\end{equation}
takes two distinct \emph{forms}, termed \emph{A} and \emph{B} \cite{epjd2017}: \emph{Form A} arises for $\eta<2\Delta\zeta$, in which case the induced dipole potential $\overline{V}_{\alpha}(\theta)$ dominates and the combined potential $\overline{V}(\theta)$ is still a double-well potential, albeit an asymmetric one, see panel (b) of Fig. \ref{fig:Comparison_Opt_Comb_1_10}. The combined potential has a global minimum of  $-\eta-\zeta_{\parallel}$ at $\theta=0^{\circ}$, a local minimum of $\eta-\zeta_{\parallel}$ at $\theta=180^{\circ}$,  and a global maximum of $\frac{\eta^2}{4\Delta\zeta}-\zeta_{\bot}$ at $\theta=\arccos\left(-\frac{\eta}{2\Delta\zeta} \right)$. Conspicuously, the members of the tunneling doublets are pushed apart and oriented due to their coupling by the permanent dipole interaction, see panels (a) and (b) of Fig. \ref{fig:Comparison_Opt_Comb_1_10}. The orientation of the two members of a given tunneling doublet thus has opposite sense, cf. the Hellmann-Feynman theorem, Eq. (\ref{H_F_2}): along the electrostatic field for the lower member that gets pushed down (this is termed \emph{right-way orientation}) and against the electrostatic field for the upper member that gets pushed up (\emph{wrong-way orientation}). This is reflected in the opposite signs of the orientation cosine shown in panel (b) of  Fig. \ref{fig:Comparison_Opt_Comb_1_10}.

On the other hand, \emph{Form B} arises for $\eta>2\Delta\zeta$, in which case $\overline{V}_{\mu}(\theta)$ dominates and $\overline{V}(\theta)$ becomes a single-well potential, with a minimum of $-\eta-\zeta_{\parallel}$ at $\theta=0^{\circ}$ and a maximum of $\eta-\zeta_{\parallel}$ at $\theta=180^{\circ}$. The states produced by the Form B combined potential are oriented and their orientation is enhanced compared with the orientation produced by the permanent dipole interaction alone at the same value of $\eta$. Although quite small for small $\Delta\zeta$, the enhancement becomes significant  at higher values of the $\Delta\zeta$ parameter, see Fig \ref{fig:Orientation}.

We note that for $\eta=2\Delta\zeta$, the potential is a single well whose maximum at $\theta=\pi$ is flat. 

The dependence of the eigenenergies of the six lowest pendular states on $\Delta\zeta$ for a fixed value of the permanent dipole interaction parameter, $\eta=10$, is shown in panel (b) of Fig. \ref{fig:Comparison_Eigenenergies_0_10}. The closer the levels in the optical field alone, cf. Fig. \ref{fig:Comparison_Eigenenergies_0_10}a, the more they are pushed apart by the superimposed permanent dipole interaction. For any given tunneling doublet and a value of $\Delta\zeta$, the levels are repelled proportionately to the value of the permanent dipole interaction parameter, $\propto \eta$. The members of the split-up tunneling doublets included in Fig. \ref{fig:Comparison_Eigenenergies_0_10}b are marked by the red vertical arrows.  

The linear scaling of the tunneling splitting with $\eta$ results in a pattern of intersections, as the pushed-up upper member of a lower tunneling doublet is bound to meet the pushed-down lower member of the upper tunneling doublet. The loci of the intersections have an analytic form: they occur at 
\begin{equation}
\label{index}
\Delta\zeta=\frac{\eta^2}{4k^2}
\end{equation}
with $k$ an integer, $k=1,2,3,...$, termed the \emph{topological index} \cite{SchmiFriJCP2014}. All the intersections are avoided as they originate from opposite parity levels coupled by the parity-mixing permanent dipole interaction. An example of such an avoided crossing is included in Fig. \ref{fig:Comparison_Eigenenergies_0_10}b and its position marked by the blue arrow. It entails the $\ket{\tilde{J}=1,|M|=0;\eta,\Delta\zeta}$ (upper member of a lower tunneling doublet) and $\ket{\tilde{J}=2,|M|=0;\eta,\Delta\zeta}$ (lower member of an upper tunneling doublet) states. For $\eta=10$, their avoided crossing occurs at $\Delta\zeta=25$ with $k=1$.

We note that the eigenproblem for a rotor subject to the combined interactions, Eq. \ref{redham}, is conditionally quasi-analytically solvable, i.e., some of its solutions can be obtained analytically at particular conditions imposed on the parameters $\eta$ and $\Delta\zeta$. Remarkably, these conditions are fulfilled at the loci of the avoided intersections \cite{SchmiFriJCP2014,SchatzFriBeckSchmiPRA2018}. For instance, for the ground state $\ket{\tilde{J}=0,|M|=0;\eta,\Delta\zeta=\eta^2/4}$, the analytic eigenenergy is $E_{0,0}=-\eta^2/4=-\Delta\zeta$ and the orientation cosine $\langle \cos \theta \rangle_{0,0}=\coth \eta-1/\eta$, cf. Ref.\cite{SchatzFriBeckSchmiPRA2018}. 

\section{The trapping potential}
\label{sec:trap_pot}

We begin by noting that the characteristic time scale for hybridizing the rotor states by the permanent or induced dipole potential  is given by the rotational period $\tau_{\mathrm{r}}$ of the molecule, as follows from the time-dependent Schr\"odinger equation \cite{OrtigosoJCP1998,CaiMarFriPRL2001}. Table \ref{table:parameters} lists the rotational periods for a sampling of linear polar molecules. On the other hand, the motion of the molecule's center of mass in a trap is given by the trapping frequency, $\omega_X$ or $\omega_Z$, see below. Given that the ratio of, say, $\nu_X=\omega_X/(2\pi)$ to the reciprocal of the rotational period, $\nu_X\,\tau_{\mathrm{r}}$, is typically on the order of $10^{-5}$, we see that the eigenstates are created much faster than the molecule can travel across the trap. This means that the \emph{eigenenergy} of the molecule in the trapping field can instantaneously adjust to the local value of the field and thus play the role of the actual \emph{trapping potential}, $U$, acting on the molecule's center of mass, i.e., on its translation. 

We begin by examining the properties of the \emph{optical trap} when the instantaneous eigenenergy of the molecule is given solely by the induced dipole potential, Eq. (\ref{Vind}). Then we move on to examine the \emph{electro-optical trap}, which is realized by superimposing a uniform (homogeneous) electrostatic field onto the optical trap -- assuming the molecule is polar and thus subject to the permanent dipole potential, Eq. (\ref{Vperm}), in addition to the induced dipole potential due to the inhomogeneous laser intensity distribution $I(X,Z)$, Eq. (\ref{intensity}). 

\subsection{Optical trap} 
\label{sec:opt}

\begin{figure}
%[htbp]
\centering
\includegraphics[width=6cm]{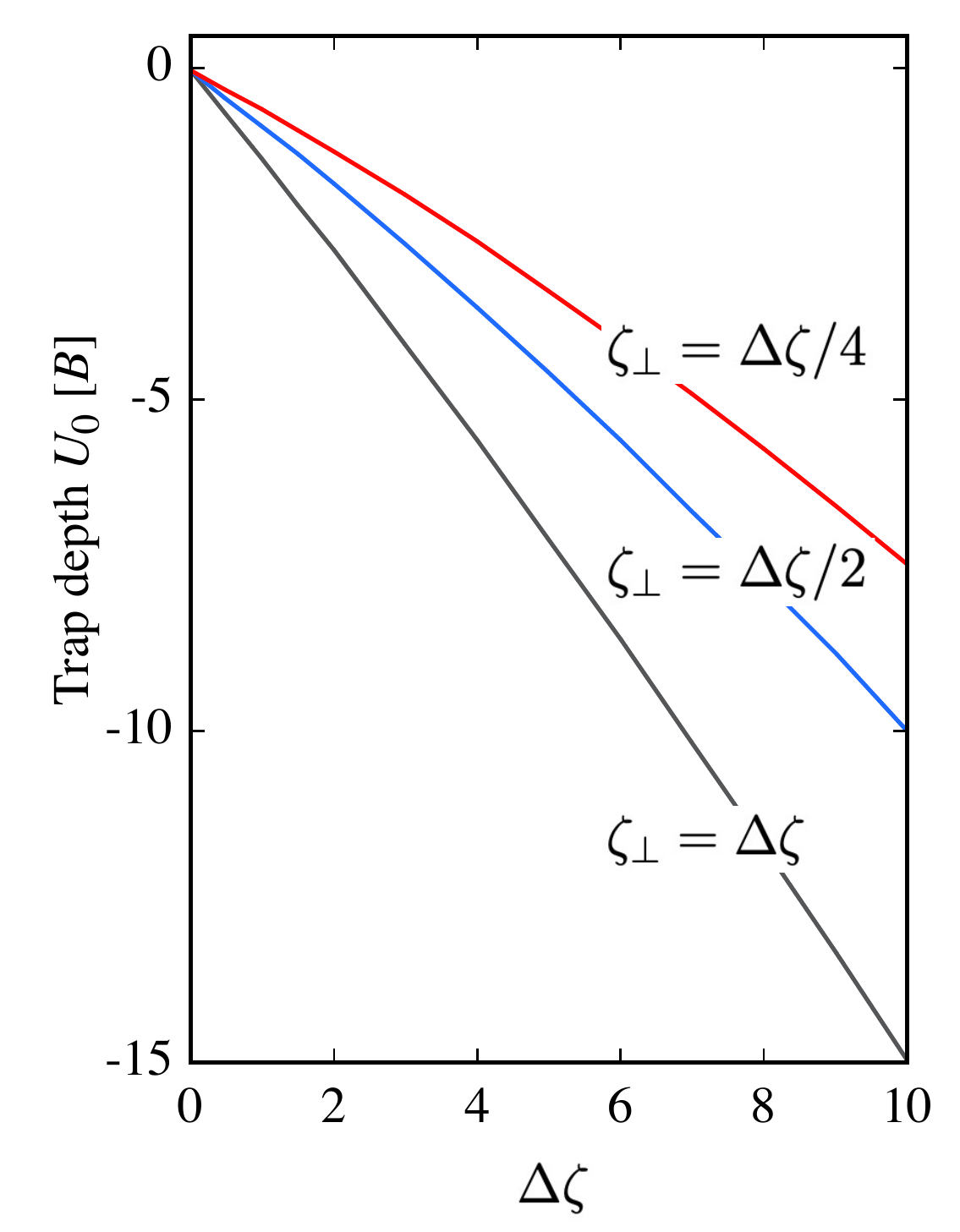}
\caption{\label{fig:TrapDepthOpt_zeta_perp} \footnotesize Dependence of the depth of the optical trap for the $\ket{\tilde{J}=0,|M|=0;\Delta\zeta}$ state on the induced dipole parameter $\Delta\zeta$ for different values of the parameter $\zeta_{\bot}$. See text.}
\end{figure}

The optical trapping potential for a linear molecule in a rotational state $\ket{\tilde{J},|M|;\Delta\zeta}$ is thus given by
\begin{equation}
\label{Ulf}
U=E_{\tilde{J},|M|}(\Delta\zeta)-B\zeta_{\bot}=E_{\tilde{J},|M|}(\Delta\zeta(X,Z))-B\zeta_{\bot}(X,Z)=U(X,Z)
\end{equation}
where $E_{\tilde{J},|M|}(\Delta\zeta(X,Z))$ is the eigenenergy of the molecule at the value of the interaction parameter 
\begin{equation}
\Delta\zeta(X,Z)=\frac{\Delta\alpha}{\epsilon_0 c B}I(X,Z)
\end{equation}
and 
\begin{equation}
\zeta_{\bot}(X,Z)=\frac{\alpha_{\bot}}{\epsilon_0 c B}I(X,Z)
\end{equation}
with $I=I(X,Z)$ the spatial distribution of the laser intensity as given by Eq. (\ref{intensity}). Fig. \ref{fig:TrapDepthOpt_zeta_perp} shows how the trap depth, $U(X=0,Z=0)\equiv U_0$, for the $\ket{\tilde{J}=0,|M|=0;\Delta\zeta}$ ground state, varies with the parameters $\Delta\zeta$ and $\zeta_{\bot}$: clearly, the larger $\zeta_{\bot}$ for a given $\Delta\zeta$, the deeper the trap. We note that $\zeta_{\bot}=\Delta\zeta/n$ implies $\zeta_{\parallel}=(n+1)\zeta_{\bot}$. Throughout this paper, we consider the case when $\zeta_{\bot}=\Delta\zeta/4$.

For weak interaction strengths, $\Delta\zeta \le 1$, the eigenenergy can be approximated by its low-field (LF) limit, Eq. (\ref{gse}). For a molecule in the rotational ground state, $\ket{\tilde{J}=0,|M|=0,\Delta\zeta}$, the optical trapping potential then takes the analytic form
\begin{equation}
\label{Ulow}
U^{(\mathrm{LF})}(X,Z)\approx -B\left(\frac{1}{3}\Delta\zeta(X,Z)+\zeta_{\bot}(X,Z)\right)=-\frac{I(X,Z)}{\epsilon_0 c}\left(\frac{1}{3}\Delta\alpha+\alpha_{\bot}\right)
\end{equation}
By making use of the \emph{mean value of the polarizability}, 
\begin{equation}
\alpha\equiv \frac{1}{3}(\alpha_{xx}+\alpha_{yy}+\alpha_{zz})=\frac{1}{3}(2\alpha_{\bot}+\alpha_{\parallel})
\end{equation} 
and of Eq. (\ref{intensity}), we can recast Eq. (\ref{Ulow}) as
\begin{equation}
\label{opticaltrap}
U^{(\mathrm{LF})}(X,Z)\approx -\frac{\alpha}{\epsilon_0 c}I(X,Z)=-\frac{U^{(\mathrm{LF})}_0}{1+\left (\frac{X}{X_R}\right)^2}\exp\left[-\frac{2Z^2}{w^2(X)}\right]
\end{equation}
with the trap depth
\begin{equation}
U^{(\mathrm{LF})}_0=\frac{\alpha}{\epsilon_0 c}I_0
\end{equation}

\subsection{Electro-optical trap}
\label{sec:el_opt}

For a trap based on the combined permanent and induced dipole interaction, the trapping potential for a molecule in a $\ket{\tilde{J},|M|;\eta,\Delta\zeta}$ state becomes
\begin{equation}
\label{E_O}
U=U(\eta,\Delta\zeta (X,Z),\zeta_{\bot}(X,Z))=E_{\tilde{J},|M|}(\eta,\Delta\zeta(X,Z))-B\zeta_{\bot}(X,Z)-E_{\tilde{J},|M|}(\eta,0)
\end{equation}
where we took into account that the induced-dipole interaction parameters have spatial distributions $\Delta\zeta=\Delta\zeta(X,Z)$ and $\zeta_{\bot}=\zeta_{\bot}(X,Z)$ given by the distribution of the laser intensity, cf. Eq. (\ref{intensity}), and that the permanent dipole interaction $\eta$ is isotropic (the electrostatic field is uniform). The second term accounts for the overall shift  due to the permanent dipole interaction of the eigenenergy by which the molecule is trapped. Note that for $\eta=0$, this term vanishes identically and we recover Eq. (\ref{Ulf}) for the optical trap. 

 \begin{figure}
%[htbp]
\centering
\includegraphics[width=6cm]{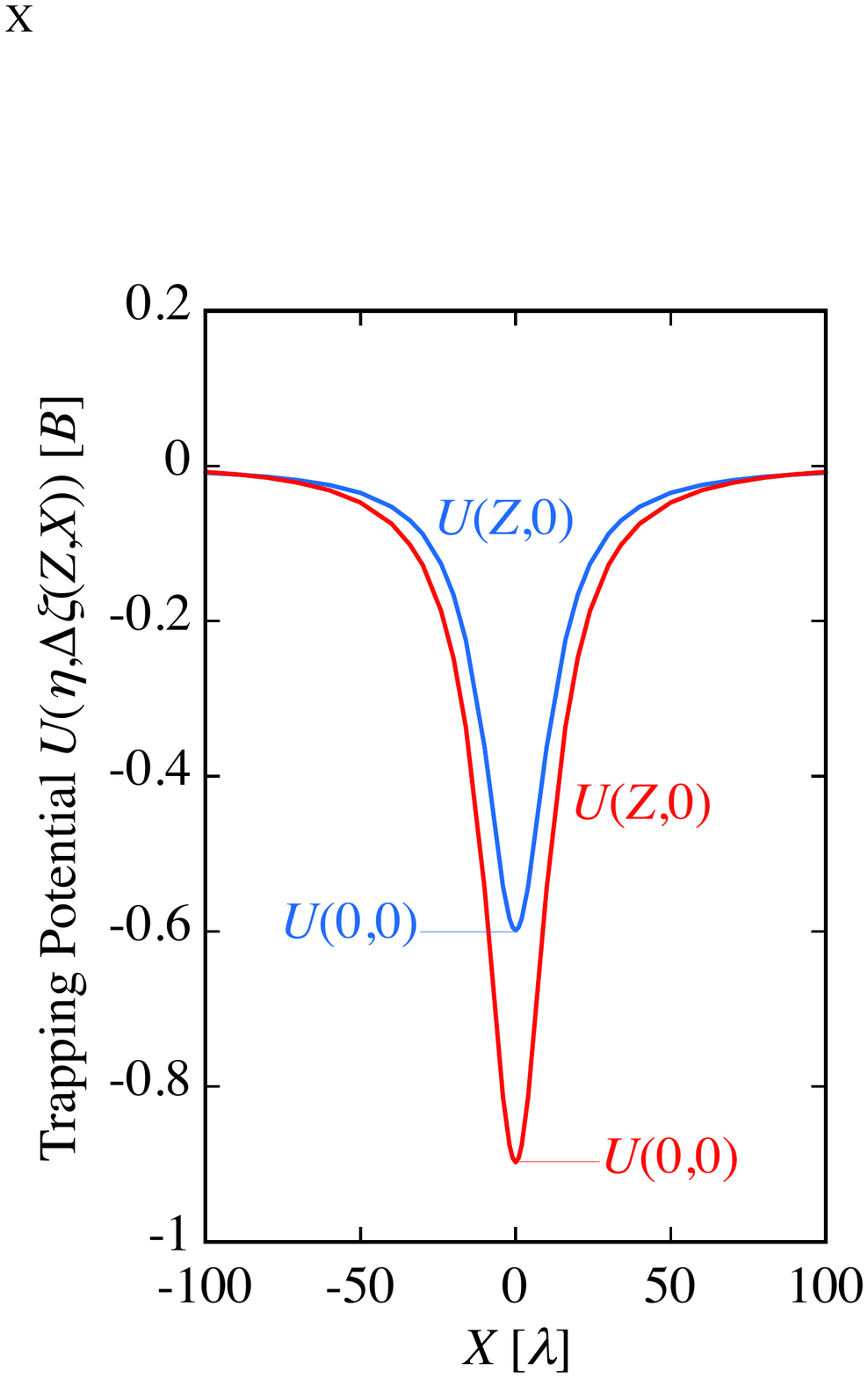}
\caption{\label{fig:Trap_CombFields_10_1_0_25} \footnotesize Comparison of the trapping potentials for a purely optical trap (blue) and for an electro-optical trap (red) for the same distribution of laser intensity $I(X,Z=0)$ and hence the same distributions of the induced-dipole interaction parameters $\Delta\zeta(X,0)$ and $\zeta_{\bot}(X,0)$ along the laser propagation direction $X$, for $\Delta\zeta(X=0,Z=0)=1$, $\zeta_{\bot}(X=0,Z=0)=\frac{1}{4}$, and $\eta=10$, cf. also Fig. \ref{fig:Comparison_Electro_10_Opt_Comb_1_10}. Note that the trapping potential is expressed in terms of the rotational constant $B$ of the trapped molecule and that the length scale of the $X$ coordinate is expressed in terms of the wavelength $\lambda$ of the optical field.}
\end{figure}

Like for the optical trap, the minimum of the trapping potential for the electro-optical trap -- its trap depth, $U_0$ -- obtains at the maximum laser intensity, $I(X=0,Z=0)\equiv I_0$, cf. Eq. (\ref{intensity}),
\begin{eqnarray}
U_0&=&U(\eta,\Delta\zeta (X=0,Z=0),\zeta_{\bot}(X=0,Z=0))\\ \nonumber &=&E_{\tilde{J},|M|}(\eta,\Delta\zeta(X=0,Z=0))-B\zeta_{\bot}(X=0,Z=0)-E_{\tilde{J},|M|}(\eta,0)
\end{eqnarray}

\begin{figure}
%[htbp]
\centering
\includegraphics[width=12cm]{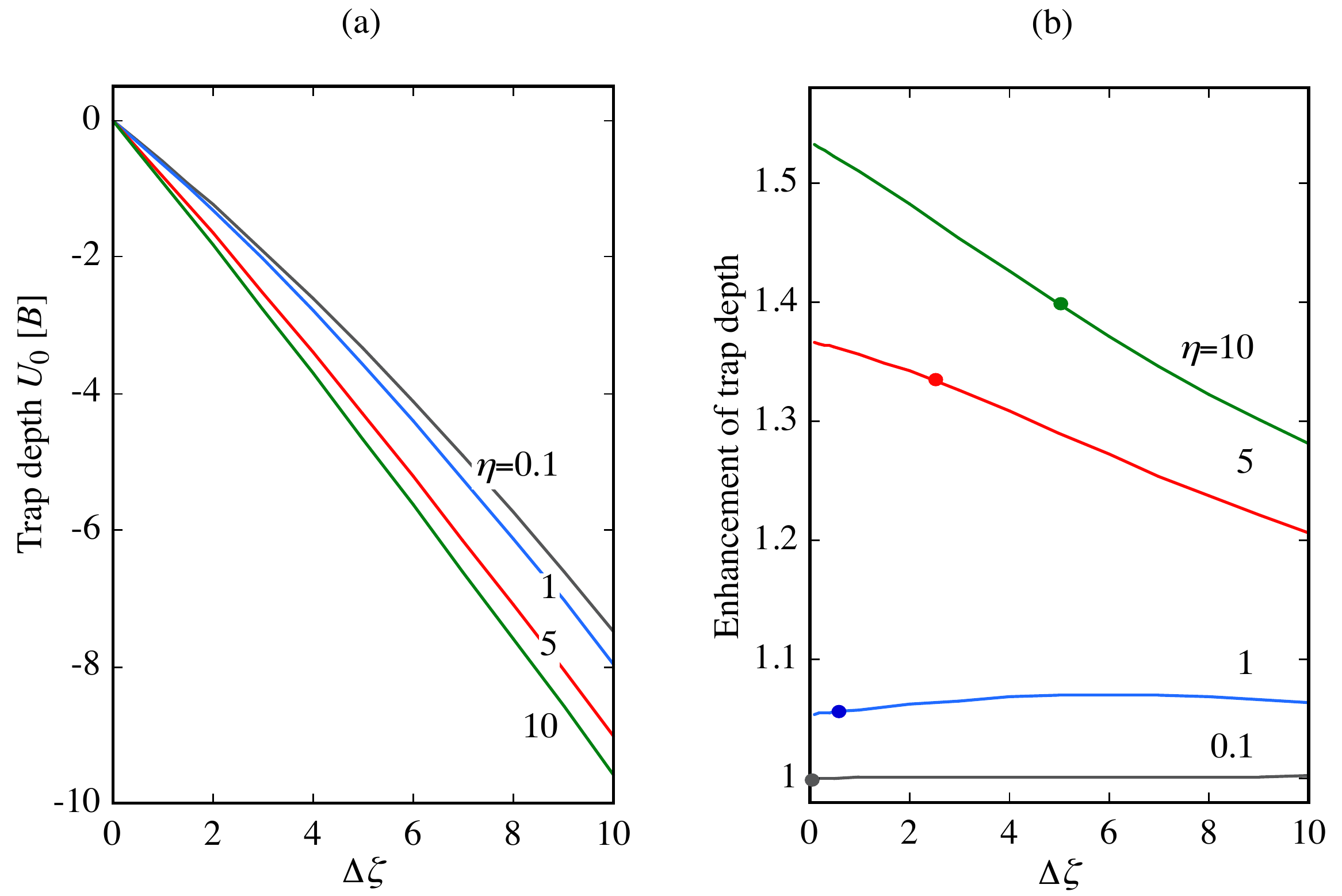}
\caption{\label{fig:TrapDepth_vs_Dzeta} \footnotesize (a) Dependence of the trap depth $U_0$ for the $\ket{\tilde{J}=0,|M|=0;\Delta\zeta,\eta}$ state on the induced dipole interaction parameter $\Delta\zeta$ for different constant values of the permanent dipole interaction parameter $\eta$.  (b) Enhancement of the trap depth compared to that of a purely optical trap as a function of the induced dipole interaction parameter $\Delta\zeta$ for different constant values of the permanent dipole interaction parameter $\eta$. The dots indicate loci where $\Delta\zeta=\eta/2$, i.e., Form A and Form B potentials are to the right and to the left of the dots, respectively. See text.}
\end{figure}

In what follows, we will consider the case when the polar and polarizable molecule is trapped via its ground state $\ket{\tilde{J}=0,|M|=0;\Delta\zeta,\eta}$. We note that in order to evaluate this state's eigenenergy -- and thus the trapping potential -- the Hamiltonian matrix has to be diagonalized ``point-by-point'' at each value of the parameters $\Delta\zeta(X,Z)$ and $\zeta_{\bot}(X,Z)$ for a given (constant) value of the parameter $\eta$. An example of such a calculation is shown in Fig. \ref{fig:Trap_CombFields_10_1_0_25} for $\eta=0$ (optical trap) and $\eta=10$ (electro-optical trap).  Clearly, the superimposed electrostatic field increases the depth of the trap, typically by 25\% to 50\%, depending on the combination of the values of the parameters involved. But how does the superimposed uniform electrostatic field amplify the inhomogeneity of the optical field as given by the spatial distribution of the Gaussian laser beam?

\begin{figure}
%[htbp]
\centering
\includegraphics[width=14cm]{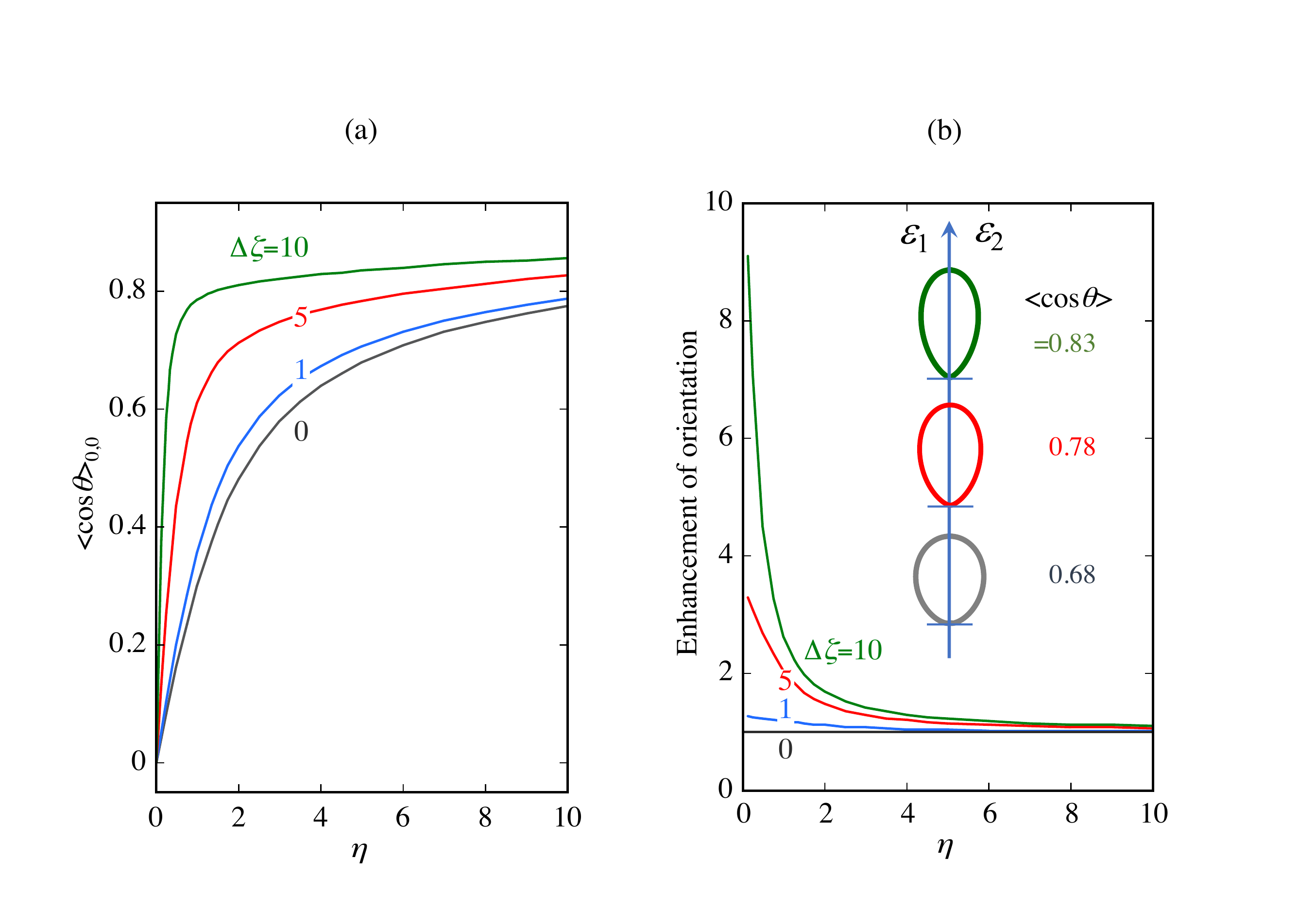}
\caption{\label{fig:Orientation} \footnotesize (a) Dependence of the orientation cosine $\langle \cos \theta \rangle_{0,0}$ of the $\ket{\tilde{J}=0,|M|=0;\Delta\zeta,\eta}$ state on the permanent dipole interaction parameter $\eta$ for different constant values of the induced dipole interaction parameter $\Delta\zeta$. (b) Enhancement of the orientation cosine $\langle \cos \theta \rangle_{0,0}$ of the $\ket{\tilde{J}=0,|M|=0;\Delta\zeta,\eta}$ state compared to that of the $\ket{\tilde{J}=0,|M|=0;\Delta\zeta=0,\eta}$ state as a function of the permanent dipole interaction parameter $\eta$ for different constant values of the induced dipole interaction parameter $\Delta\zeta$. Also shown are polar plots of the squares of the corresponding wavefunctions for $\eta=5$ and different values of $\Delta\zeta$ whose color-coding is the same as that of the labeled curves. See text.}
\end{figure}

A clue as to why this amplification takes place is provided by Fig. \ref{fig:TrapDepth_vs_Dzeta} whose panel (a) shows the dependence of the trap depth $U_0$ on the induced-dipole parameter $\Delta\zeta$ for different fixed values of the permanent dipole parameter $\eta$: for a given value of $\eta$, the greater $\Delta\zeta$, the deeper the trap. Panel (b) shows the enhancement of the trap depth as a ratio of $U_0(\Delta\zeta,\eta)/U_0(\Delta\zeta,\eta=0)$, i.e., as a depth of the electro-optical trap relative to the depth of the purely optical trap. Thus when the trap depth of the optical trap (blue) in Fig. \ref{fig:Trap_CombFields_10_1_0_25} is at its maximum, so is the enhancement of the trap depth by the superimposed permanent dipole interaction (red). 

There are  two quantum mechanisms involved in enhancing the trap depth, depending on whether \emph{Form A} or \emph{Form B} potential is at play, cf. Sec. \ref{sec:combined} and Fig. \ref{fig:TrapDepth_vs_Dzeta}b, which shows the loci where Form A potential morphs into Form B potential. Apparently, the transition between the two forms as reflected in the trap depth enhancement is quite smooth. For Form A (double-well potential dominated by the induced-dipole interaction), the $\ket{\tilde{J}=0,|M|=0;\Delta\zeta,\eta}$ state is the lower member of a tunneling doublet (the upper member doesn't have to be bound by the combined potential) and, therefore, is pushed down as it is coupled to the upper doublet member by the permanent dipole interaction. This coupling -- and thus the downward push of the energy level -- is the stronger the greater the value of the induced-dipole interaction parameter $\Delta\zeta$. On the other hand, the enhancement of the trap depth for the Form B potential (single well dominated by the permanent dipole interaction) -- relevant to what we see in Fig. \ref{fig:Trap_CombFields_10_1_0_25} -- can be explained by the increased right-way orientation (i.e., along the electrostatic field vector $\boldsymbol{\varepsilon}_1$) of the $\ket{\tilde{J}=0,|M|=0;\Delta\zeta,\eta}$ state and the corresponding downward shift of its eigenenergy as ordained by the Hellmann-Feynman theorem, cf. Eq. \ref{H_F_2}. The increase in the orientation cosine $\langle \cos \theta \rangle_{0,0}$ of the $\ket{\tilde{J}=0,|M|=0;\Delta\zeta,\eta}$ state at a given $\eta$ with $\Delta\zeta$ is illustrated in Fig. \ref{fig:Orientation}. It arises, in turn, from an increased confinement of the librational amplitude of the molecular axis by the induced dipole interaction. While panel (a) shows the effects of the induced-dipole interaction on the orientation cosine, panel (b) displays the enhancement factor defined as the ratio of the orientation cosine with the optical field on to the orientation cosine in the absence of the optical field. Also shown are polar plots of the squares of the wavefunctions of the $\ket{\tilde{J}=0,|M|=0;\Delta\zeta,\eta}$ state at $\eta=5$ and increasing values of $\Delta\zeta$, which attest to the ever narrower angular confinement of the molecular axis with $\Delta\zeta$. 

Thus we see that the synergy between the permanent and induced dipole interactions enhances the trap depth that would obtain for the optical field alone while, at the same time, increasing the orientation of the trapped molecule beyond what it would be in the electrostatic field alone.  

\subsection{Harmonic electro-optical trap} 
\label{sec:harm}

A power-series expansion of the laser intensity around $X,Z = 0$ up to the 2$^{\mathrm{nd}}$ order approximates the laser intensity at the center of the trap as
\begin{eqnarray}
I(X\rightarrow 0,Z \rightarrow 0) & \approx & I_0\left[1-\frac{X^2}{X_{\mathrm{R}}^2}+\frac{4I_0X^2Z^2}{w_0^2 Z_{\mathrm{R}}^2}-\frac{2Z^2}{w_0^2}\right]\\ \nonumber & \approx & I_0\left[1-\left(\frac{X}{X_{\mathrm{R}}}\right)^2-2\left(\frac{Z}{w_0}\right)^2 \right]
\end{eqnarray}

\begin{figure}
%[htbp]
\centering
\includegraphics[width=6cm]{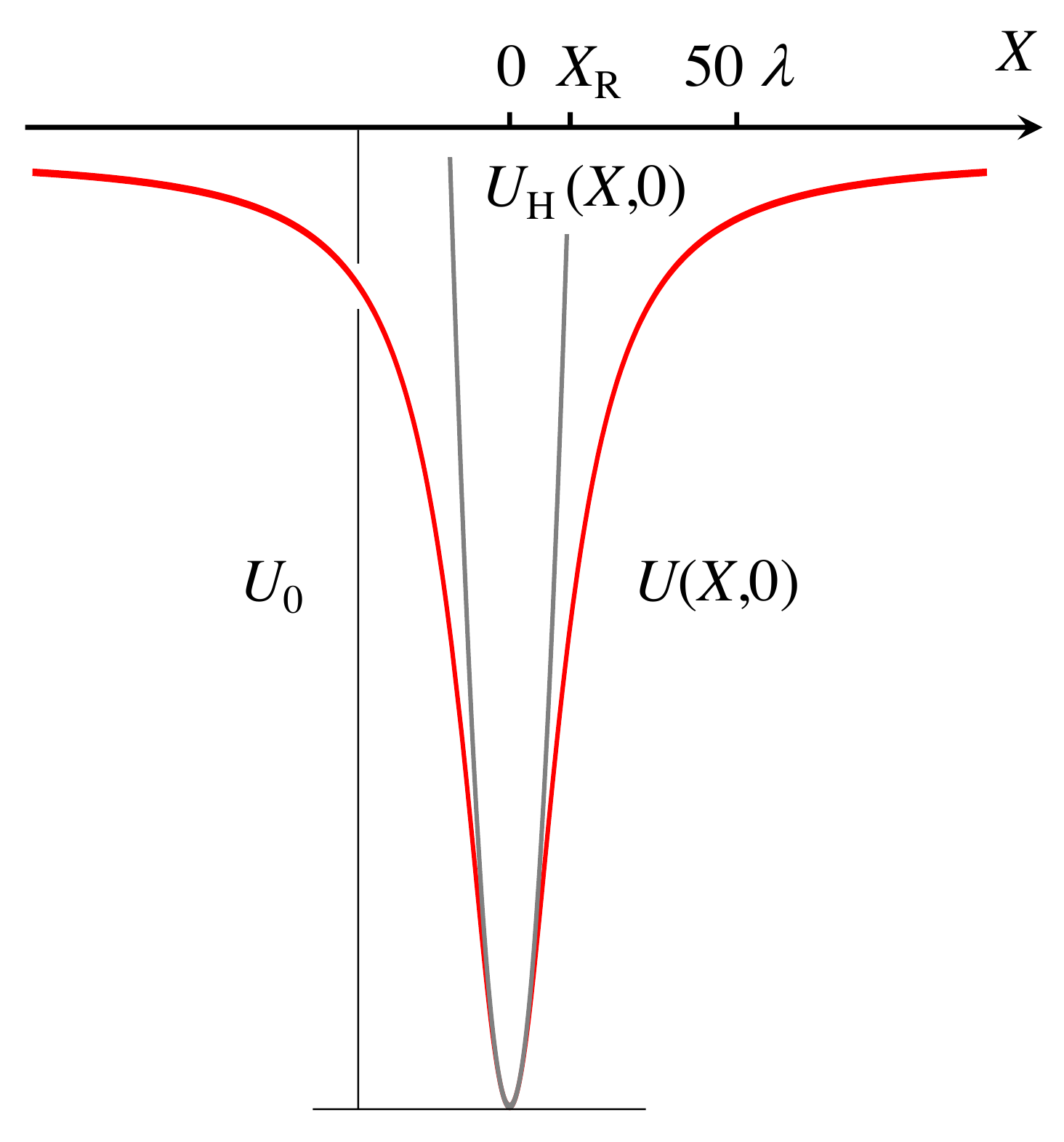}
\caption{\label{fig:Trap_and_H} \footnotesize The trapping potential $U(X,Z)$ of Eq. (\ref{E_O}) shown in red together with its harmonic counterpart $U_{\mathrm{H}}(X,Z)$ of Eq. (\ref{H}) shown in gray, plotted for $Z=0$.  Note that the harmonic trapping potential approximates  the electro-optical trapping potential faithfully up to about the Rayleigh length $X_{\mathrm{R}}$.  }
\end{figure}

The harmonic trapping potential
\begin{equation}
\label{H}
U_{\mathrm{H}}=-|U_0|\left[1-\left(\frac{X}{X_{\mathrm{R}}}\right)^2-2\left(\frac{Z}{w_0}\right)^2 \right]
\end{equation}
is shown together with the trapping potential in Fig. \ref{fig:Trap_and_H}. It approximates well 
the trapping potential up to the Rayleigh length $X_{\mathrm{R}}$.

The characteristic trapping frequencies of the  \emph{harmonic electro-optical trap} of Eq. (\ref{H})
obtain by equating the mutually corresponding terms of the harmonic oscillator potential,
\begin{equation}
|U_0|\left(\frac{X}{X_{\mathrm{R}}}\right)^2=\frac{1}{2}m\omega_X^2 X^2
\end{equation}
\begin{equation}
2|U_0|\left(\frac{Z}{w_0}\right)^2=\frac{1}{2}m\omega_Z^2 Z^2
\end{equation}
yielding
\begin{eqnarray}
\label{trapfrequencies}
\omega_X=\left(\frac{2|U_0|}{mX_{\mathrm{R}}^2}\right)^{1/2} & \hspace{0.5cm} \mathrm{and} & \hspace{0.5cm} \omega_Z=\left(\frac{4|U_0|}{mw_0^2}\right)^{1/2}
\end{eqnarray}
with  $m$ the mass of the molecule.

The trapping frequencies, Eq. (\ref{trapfrequencies}), make it possible to evaluate  the root-mean-square velocity, $v_{\mathrm{rms}}$, of the molecules confined by the harmonic optical trap.   For instance, for the $X$ direction (along the laser beam), this is 
\begin{equation}
 v_{\mathrm{rms},X}=\left(\frac{\hbar \omega_X}{m}\right)^{1/2}
 \end{equation}
For a time-of-flight expansion over time $t$ of the molecular cloud released from the trap, we then obtain
\begin{equation}
X(t)^2=X(0)^2+v^2_{\mathrm{rms},X}t^2=X(0)^2\left(1+\omega^2_{X(t)}t^2\right)
\end{equation}
Determining the expansion from an initial value $X(0)$ to a final value $X(t)$ then gives the temperature of the cloud in the $X$ direction,
\begin{equation}
T=\frac{1}{2k_\mathrm{B}}m\omega_X^2 X(0)^2=\frac{1}{2k_B}m\omega_X^2 \left(\frac{X(t)^2}{1+\omega_X^2 t^2}\right)
\end{equation}
where $k_{\mathrm{B}}$ is Boltzmann's constant.

\subsection{Trapping of polar molecules in perpendicular optical and electrostatic fields} 
\label{sec:perp}

\begin{figure}
%[htbp]
\centering
\includegraphics[width=12cm]{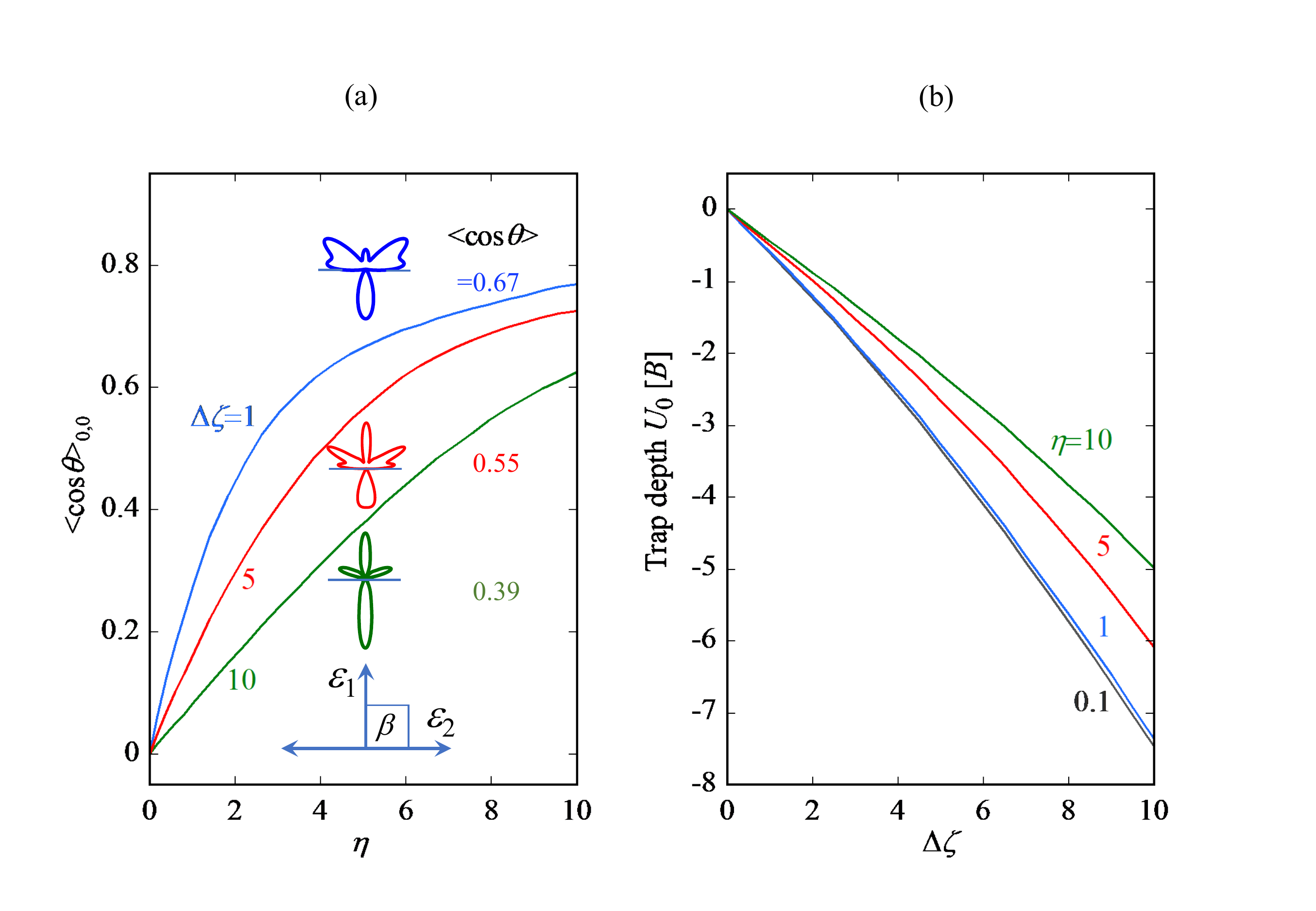}
\caption{\label{fig:PerpendicularFields} \footnotesize (a) Dependence of the orientation cosine $\langle \cos \theta \rangle_{0,0}$ of the $\ket{\tilde{J}=0,\tilde{M}=0;\Delta\zeta,\eta}$ state on the permanent dipole interaction parameter $\eta$ for different constant values of the induced dipole interaction parameter $\Delta\zeta$ for perpendicular fields $\boldsymbol{\varepsilon}_1$ and $\boldsymbol{\varepsilon}_2$ (tilt angle $\beta=\pi/2$).  Also shown are polar plots of the squares of the corresponding wavefunctions for $\eta=5$ and different values of $\Delta\zeta$ whose color-coding is the same as that of the labeled curves. The horizontal bars run through the origin of the polar coordinate system and divide the right-way (upward) and wrong-way (downward) oriented lobes of the wavefunctions. (b) Dependence of the trap depth $U_0$ for the $\ket{\tilde{J}=0,\tilde{M}=0;\Delta\zeta,\eta}$ state in perpendicular fields on the induced dipole interaction parameter $\Delta\zeta$ for different constant values of the permanent dipole interaction parameter $\eta$. See text.}
\end{figure}

In non-collinear, tilted fields, when the $\boldsymbol{\varepsilon}_1$ and $\boldsymbol{\varepsilon}_2$ vectors make an angle $\beta \ne 0,\pi$, the two fields compete with one another and their effects are no longer synergistic as each field forces the dipole to disfavor the  direction of the other field \cite{FriHerJCP1999,FriHerJPC1999,HartFriJCP2008}. Maximum competition arises for perpendicular fields, $\beta=\pi/2$, when an increased induced-dipole interaction suppresses the molecule's orientation along the electrostatic field, cf. panel (a) of Fig. \ref{fig:PerpendicularFields}. In addition, the competition between the tilted fields causes the azimuthal angles of the molecular axis about the two field vectors to be nonuniformly distributed. However, the molecular axis remains symmetrically distributed with respect to the plane defined by the two field vectors and for perpendicular fields, the problem has a $C_{2v}$ symmetry. As noted, in Section \ref{sec:eigenproblem}, $M$ is no longer a good quantum number in tilted fields but can serve, along with $\tilde{J}$,  as an adiabatic label, $\tilde{M}$, of a given field-dressed state:  $\ket{\tilde{J},\tilde{M};\eta=0,\Delta \zeta=0}\rightarrow \ket{J,|M|}$.

The competition between the perpendicular fields also transpires in the shape of the corresponding eigenfunctions, cf. the polar plots of the eigenfunctions squared in Fig. \ref{fig:PerpendicularFields}a. Whereas in the absence of the electrostatic field, the wavefunction has the shape of a horizontal p-orbital (for a horizontal polarization of the optical field), turning on the permanent dipole interaction in the vertical direction adds new lobes. The proportions (relative surface areas) of the lobes vary with the values of the interaction parameters.

As illustrated in panel (b) of Fig. \ref{fig:PerpendicularFields}, adding a perpendicular electrostatic field diminishes the trap depth due to the optical field and does the more so the greater the strength of the permanent dipole interaction. This contrasts with the effect of a collinear electrostatic field that enhances the trap depth. 

Given the opposite effects of collinear and perpendicular fields on the trap depth and the orientation of the trapped molecules, one could significantly -- and quickly -- alter either by tilting the polarization plane of the optical field with respect to the electrostatic field. Thereby an electro-optical trap offers yet another element of control of the confined molecules. 

\section{Conclusions and Prospects}
\label{sec:concl}

The quantum treatment of optical traps -- or tweezers -- for molecules presented herein provides a detailed recipe for  designing a trap with preordained effects on both the translation and rotation of the molecules to be trapped. These effects depend on the dimensionless parameter $\Delta\zeta$ that reflects, apart from the intensity of the optical field, the anisotropic polarizability and the moment of inertia/rotational constant of the molecules. Only in the low-field limit, $\Delta\zeta \rightarrow 0$, and for the ground initial rotational state of the molecules, does the eigenenergy -- and thus the trap depth -- scale with the average molecular polarizability and the rotational constant while the alignment imparted to the molecules remains puny. In all other cases, the anisotropy of the molecular polarizability has to be taken into account and for interaction strengths such that $\Delta\zeta \ge 1$ (corresponding to an intensity of the laser field of $10^{7}$ W/cm$^2$ or greater for a number of  ``popular'' molecules, cf. Table \ref{table:parameters}), the pendular eigenproperties that define the trap have to be calculated by solving the eigenproblem for a rigid rotor subject to the induced electric dipole interaction, see Appendix \ref{sec:appendix2}. 

However, the main focus of the present paper is on the electro-optical trap/tweezer, which is realized by embedding an optical trap in a collinear uniform electrostatic field. The effects of the electro-optical trap on molecular translation and rotation depend on a pair of dimensionless parameters, $\Delta\zeta$ and $\eta$. While the $\Delta\zeta$ parameter retains its original meaning, the $\eta$ parameter takes into account the body-fixed electric dipole moment and the moment of inertia of the molecules to be confined. There is no low-field limit expression available for calculating the trap depth or the directionality -- both alignment and orientation -- of the pendular states of the molecules confined by the electro-optical trap and so the trap's effects have to be evaluated by solving the eigenproblem for a rigid rotor subject to a combined permanent and induced electric dipole interaction, see Appendix \ref{sec:appendix2}. However, analytic solutions exist for particular ratios of the interaction parameters $\eta$ and $\Delta\zeta$ corresponding to integer values of the topological index $k$, cf. Eq. (\ref{index}).

Although the optical trap/tweezer obtains as a special case of the electro-optical trap for $\eta=0$, the combined interaction amounts to more than a sum of its parts. In the context of molecular trapping, this shows in enhancing the trap depth due to the optical field alone and the orientation due to the electrostatic field alone. Both enhancement effects are quite subtle and have to do with the synergy of the two collinear permanent and induced dipole interactions that derives from their distinct eigenenergy level structures and the avoided crossings that arise from their combination. The enhancement effects are illustrated in dedicated figures (Figs. \ref{fig:TrapDepth_vs_Dzeta}b and \ref{fig:Orientation}b) as well as in a generic plot of the trap depth, Fig. \ref{fig:Trap_CombFields_10_1_0_25}. 

Apart from enhancing the trap depth and lending orientation to the trapped polar molecules, the electro-optical trap offers the possibility to lift the degeneracy of the $\pm M$ levels as well as to rapidly vary both the orientation and trap depth by tilting the polarization plane of the optical field with respect to the electrostatic field (Fig. \ref{fig:PerpendicularFields}). 

Thus electro-optical trapping via a certain oriented state of a polar polarizable molecule amounts to state preparation of this particular directional state. The electro-optical trap/tweezer may, therefore, facilitate some of the applications of molecular trapping mentioned in the Introduction, especially of quantum computing and simulation \cite{Wei_arXiv2022} and detailed collision stereodynamics \cite{Dulieu_Oster_RSC_2018} that distinguishes between heads versus tails in molecular encounters. The added electrostatic field may also decouple hyperfine levels and thereby prolong the rotational coherence times achieved so far in an optical field alone \cite{KettNiDoylePRL2021}. The beneficial effect of a superimposed electrostatic field onto an optical trap has in fact been already demonstrated as a means to enhance the dipolar evaporative cooling rate by making tunable dipolar interactions dominate over all inelastic processes \cite{Ye_ElectricFieldShilding_Science_2020,Ye_DipolarEvap_Nature_2020}. Alternatively, the synergistic enhancement of the trap depth of an electro-optical trap affords a reduced intensity of the optical field which would foster microwave shielding of the elastic channel and thereby the elastic-to-inelastic collision rate \cite{NiKettDoyleScience2021}. 
\newpage

\section{Appendices}

\subsection{Permanent and induced-dipole potentials}
\label{sec:appendix1}

\begin{figure}
%[htbp]
\centering
\includegraphics[width=10cm]{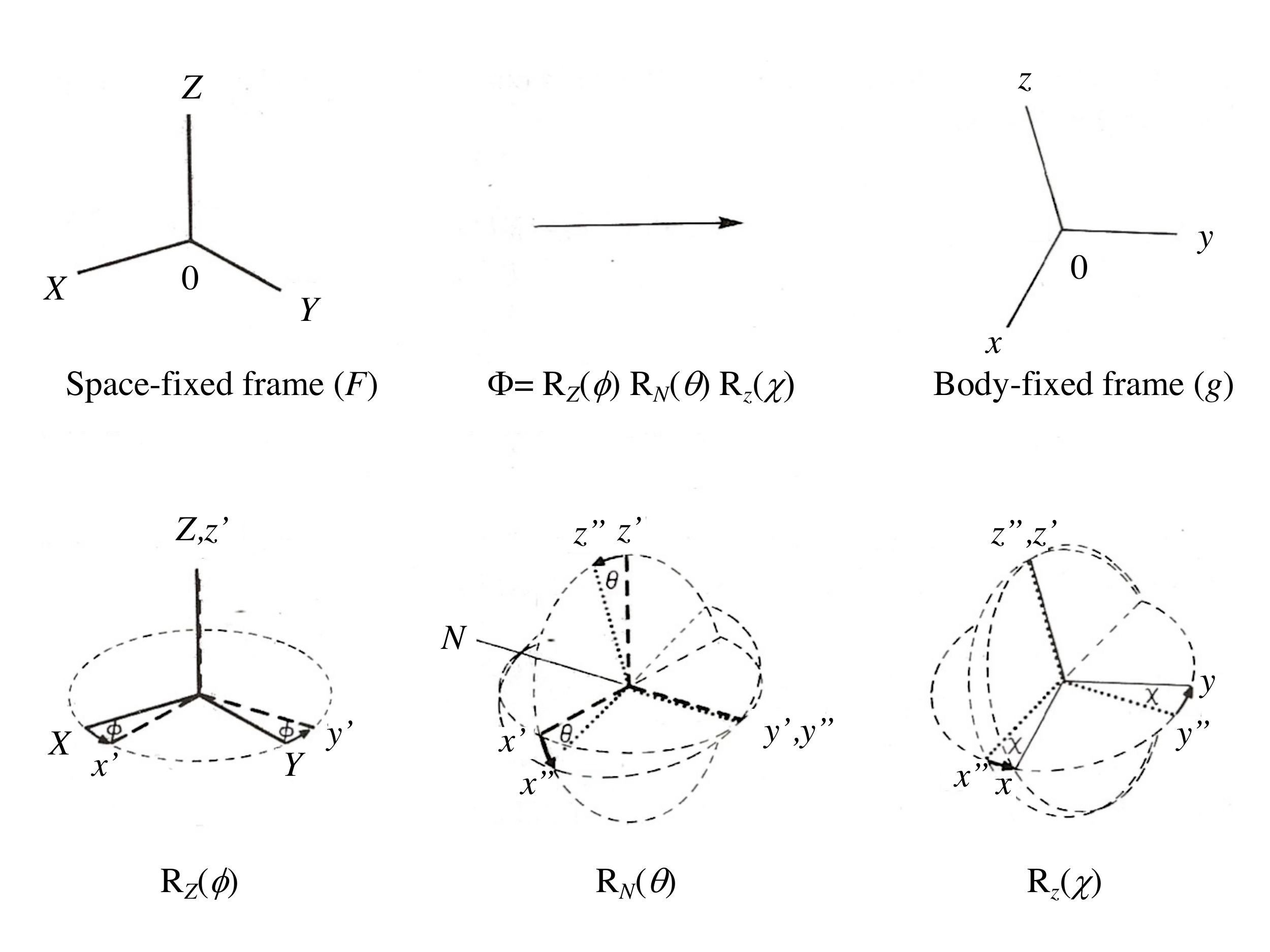}
\caption{\label{fig:frame} \footnotesize Transformation from the space-fixed frame $(X,Y,Z)$ to the body-fixed frame $(x,y,z)$, effected by the inverse direction cosine matrix $\Phi$. Also shown is the parametrization of the transformation by the Euler angles $\phi$, $\theta$, and $\chi$. Adapted from Ref. \cite{Zare}.}
\end{figure}

The transformation from the body-fixed frame $(x,y,z)$ to the space-fixed frame $(X,Y,Z)$, see Fig. \ref{fig:frame}, is effected by the direction cosine matrix, $\Phi$, given by
\begin{equation}
\Phi = 
\begin{pmatrix}
Xx & Xy & Xz \\
Yx & Yy & Yz \\
Zx & Zy & Zz 
\end{pmatrix}
\end{equation}
where
\begin{eqnarray}
Xx=&\cos\phi \cos\theta \cos\chi-\sin\phi \sin\chi \\ \nonumber
Xy=&-\cos\phi \cos\theta \sin\chi-\sin\phi \cos\chi \\ \nonumber
Xz=&\cos\phi \sin\theta \\ \nonumber
\medskip
Yx=&\sin\phi \cos\theta \cos\chi+\cos\phi \sin\chi \\ \nonumber
Yy=&-\sin\phi \cos\theta \sin\chi-\cos\phi \cos\chi \\ \nonumber
Yz=&\sin\phi \sin\theta \\ \nonumber
\medskip
Zx=&-\sin\theta \sin\chi \\ \nonumber
Zy=&\sin\theta \sin\chi \\ \nonumber
Zz=&\cos\theta
\end{eqnarray}

Assuming that the space-fixed electric field has only a $Z$ component of magnitude $\varepsilon$ in the space-fixed frame, $\varepsilon_F=(0,0,\varepsilon_1)$, and that the body-fixed permanent dipole moment has only a $z$-component of magnitude $\mu$ in the body-fixed frame, $\mu_g=(0,0,\mu)$.\footnote{Herein, all products involving vectors, tensors, and matrices are dot products.} Then
\begin{eqnarray}
\label{muexample1}
\mu_g \varepsilon_g=\Phi^{-1} \mu_F\Phi^{-1} \varepsilon_F=
\Phi^{-1}\Phi \mu_g \Phi^{-1} \varepsilon_F=\mu_g \Phi^{-1}\varepsilon_F=\mu \varepsilon_1 \cos\theta
\end{eqnarray}
and
\begin{eqnarray}
\label{muexample2}
\mu_F \varepsilon_F=\Phi \mu_g \varepsilon_F=\mu \varepsilon_1 \cos\theta
\end{eqnarray}
Thus, the  potential energy of the interaction of the permanent body-fixed dipole $\mu_g$ with a space-fixed electric field $\varepsilon_F$ is given by
\begin{equation}
\label{Vmu}
V_{\mu}(\theta)=-\mu_g \varepsilon_g=-\mu_F \varepsilon_F=-\mu \varepsilon_1 \cos\theta
\end{equation}

 For the induced dipole interaction $V_{\alpha}(\theta)$ due to the electric field $\varepsilon_F$ acting on the induced dipole $\tilde{\mu}_F$ produced by the very same field acting on the molecular polarizability $\alpha_g$, we have, in the space-fixed frame,
 \begin{equation}
\label{Valpha}
V_{\alpha}(\theta)=-\tilde{\mu}_F \varepsilon_F=-\Phi \tilde{\mu}_g \varepsilon_F=-\Phi \alpha_g \varepsilon_g \varepsilon_F=-\Phi \alpha_g \Phi^{-1} \varepsilon_F \varepsilon_F
\end{equation}
with $\tilde{\mu}_g=\alpha_g \varepsilon_g$ the body-fixed induced dipole moment and $\alpha_g$ the body-fixed polarizability tensor. This second-order Cartesian tensor can be diagonalized and represented by its principal components (components along the principal body-fixed axes $x,y,z$) 
\begin{equation}
\label{principal}
\alpha_g=\left(
\begin{matrix}
    \alpha_{xx} & 0 & 0\\
    0 & \alpha_{yy} & 0\\
    0 & 0 & \alpha_{zz}
  \end{matrix}\right)
\end{equation}
Moreover, for a linear molecule, $\alpha_{xx}=\alpha_{yy}\equiv \alpha_{\perp}<\alpha_{zz}\equiv \alpha_{\parallel}$. By substituting $\alpha_g$ from Eq. (\ref{principal}) into Eq. (\ref{Valpha}) and keeping in mind that $\varepsilon_F=(0,0,\varepsilon_2)$, we obtain for the induced-dipole potential
\begin{equation}
\label{Valpha2}
V_{\alpha}(\theta)=-\varepsilon_2^2\left(\alpha_{\parallel}\cos^2\theta+\alpha_{\perp}\cos^2\theta\right)=-\varepsilon_2^2\left[(\alpha_{\parallel}-\alpha_{\perp})\cos^2\theta+\alpha_{\perp}\right]
\end{equation}
which we write as
\begin{equation}
\label{Valpha3}
V_{\alpha}(\theta)=-\varepsilon_2^2\left(\Delta\alpha \cos^2\theta+\alpha_{\perp}\right)
\end{equation}
by setting $\Delta\alpha \equiv \alpha_{\parallel}-\alpha_{\perp}$.

\newpage
\subsection{The matrix elements of the Hamiltonian of a polar and polarizable rotor subject to combined permanent and induced dipole interactions}
\label{sec:appendix2}

The matrix elements of the Hamiltonian of a polar and polarizable rotor subject to combined permanent and induced dipole interactions characterized, respectively, by the dimensionless parameters $\eta$ and $\Delta\zeta$ in the free-rotor basis set $\ket{J,|M|}$ is given by:
\begin{eqnarray}
\langle j',m'|H/B|j,m \rangle = \delta_{jj'}\delta_{mm'}[j(j+1)- \zeta_{\bot}\\ \nonumber
-\eta\cos\beta (2j+1)^{1/2}(2j'+1)^{1/2}(-1)^m 
\times 
\begin{pmatrix}j & 1 & j' \\
	-m & 0 & m\\
\end{pmatrix}
\begin{pmatrix}j & 1 & j' \\
	0 & 0 & 0\\
\end{pmatrix}\\ \nonumber
-\eta \sin\beta (2j+1)^{1/2}(2j'+1)^{1/2}(-1)^m (1/2)^{1/2}\\ \nonumber \times \left[ \begin{pmatrix}j & 1 & j' \\
	-m & -1 & m'\\ 
\end{pmatrix}
\begin{pmatrix}j & 1 & j' \\
	0 & 0 & 0\\
\end{pmatrix}-\begin{pmatrix}j & 1 & j' \\
	-m & 1 & m'\\
\end{pmatrix}
\begin{pmatrix}j & 1 & j' \\
	0 & 0 & 0\\
\end{pmatrix}\right] \\ \nonumber
-\Delta\zeta (2j+1)^{1/2}(2j'+1)^{1/2}(-1)^m 
\begin{pmatrix}j & 2 & j' \\
	-m & 0 & m\\
\end{pmatrix}
\begin{pmatrix}j & 2 & j' \\
	0 & 0 & 0\\
\end{pmatrix}\\ \nonumber
\end{eqnarray}
where $\delta_{x,x'}$ is Kronecker's delta and $\beta$ is the tilt angle between the electrostatic field vector $\boldsymbol{\varepsilon}_1$ and the optical field vector $\boldsymbol{\varepsilon}_{2}$.

In the calculations presented herein, the Hamiltonian matrix was truncated at $J_{\mathrm{max}}=60$, sufficient to achieve convergence within 0.1\% for all states and field strengths considered.

\bigskip

\noindent {\bf Acknowledgments}

I thank Mike Tarbutt (Imperial College London), Stefan Truppe (Fritz-Haber-Institut der Max-Planck-Gesellschaft, Berlin),  and Sean Burchesky (Harvard) for insightful comments, John Doyle (Harvard), Kang-Kuen Ni (Harvard), Hossein Sadeghpour (Harvard), Burkhard Schmidt (Freie Universit\"at Berlin), and Jun Ye (JILA) for a critical reading of the manuscript, and to Ben Augenbraun (Harvard), Zack Lasner (Harvard), Lan Cheng (Johns Hopkins University), and Steve Coy (MIT) for discussions about the molecular parameters involved. I greatly appreciate the hospitality of John Doyle and Hossein Sadeghpour during my stay at Harvard Physics and at the Harvard \& Smithsonian Institute for Theoretical Atomic, Molecular, and Optical Physics (ITAMP).

\medskip

\emph{This paper is dedicated to Gerard Meijer on the occasion of his 60th birthday and to Dudley Herschbach on the occasion of his 90th birthday.}

\end{document}